\documentclass[10pt, a4paper]{article}
\usepackage[margin=1in]{geometry}
\usepackage{amsthm, amsmath, amssymb}
\usepackage{array}
\usepackage{graphicx}
\usepackage{algorithm}
\usepackage[noend]{algpseudocode}
\usepackage[sort&compress,numbers]{natbib}   
\usepackage[hyphens]{url} 

\newcommand{\expnumber}[2]{{#1}\mathrm{e}{#2}}

\def\esim{\overset{e}{\sim}}
\def\dsim{\overset{d}{\sim}}

\title{Multi-modality imaging \\ with structure-promoting regularisers}
\author{Matthias J. Ehrhardt}
\date{\footnotesize Institute for Mathematical Innovation, University of Bath, UK\\ \url{m.ehrhardt@bath.ac.uk}}

\def\dataxray{\texttt{x-ray}}
\def\datasuper{\texttt{super-resolution}}

\def\TV{\operatorname{TV}}
\def\wTV{\operatorname{wTV}}
\def\dTV{\operatorname{dTV}}
\def\TGV{\operatorname{TGV}}
\def\wTGV{\operatorname{wTGV}}
\def\dTGV{\operatorname{dTGV}}
\def\JTV{\operatorname{JTV}}
\def\TNV{\operatorname{TNV}}
\def\Hone{\operatorname{H^1}}
\def\wHone{\operatorname{wH^1}}
\def\dHone{\operatorname{dH^1}}

\newcolumntype{P}[1]{>{\centering\arraybackslash}p{#1}}
\def\MySpace{\hspace*{1.5mm}}

\newcommand{\runinhead}[1]{\paragraph{#1}}
\def\svhline{\hline}

\newtheorem{definition}{Definition}
\newtheorem{remark}{Remark}

\begin{document}
\maketitle

\begin{abstract}
Imaging with multiple modalities or multiple channels is becoming increasingly important for our modern society. A key tool for understanding and early diagnosis of cancer and dementia is PET-MR, a combined positron emission tomography and magnetic resonance imaging scanner which can simultaneously acquire functional and anatomical data. Similarly in remote sensing, while hyperspectral sensors may allow to characterise and distinguish materials, digital cameras offer high spatial resolution to delineate objects. In both of these examples, the imaging modalities can be considered individually or jointly. In this chapter we discuss mathematical approaches which allow to combine information from several imaging modalities so that multi-modality imaging can be more than just the sum of its components.
\end{abstract}

\section{Introduction} \label{sec:intro}
Many tasks in almost all scientific fields can be posed as an inverse problem of the form 
\begin{equation}
K u = f \label{eq:ip}
\end{equation}
where $K$ is a mathematical model that connects an unknown quantity of interest $u$ to measured data $f$. The task is to recover $u$ from data $f$ under the model $K$. In practice this task is difficult because of measurement errors in the data $f$ and inaccuracies in the model $K$. Moreover, in many cases the model \eqref{eq:ip} lacks information we have at hand about the unknown quantity $u$ such as its regularity. In this chapter we are interested in the situation when have a-priori knowledge about the "structure" of $u$ from a second measurement $v$ which we want to exploit in the inversion. Throughout this chapter we will refer to $v$ as the \emph{side information}. Intuitively, this is the case when $u$ and $v$ describe different properties of the same geometry (in medicine: anatomy). We will be more precise in Section~\ref{section:similarity} where we discuss mathematical models for structural similarity. The two notions we will discuss in detail that the edges of the two images $u$ and $v$ having similar 1) locations~\cite{Arridge2008, Bresson2008, Haber2013, Knoll2014, Ehrhardt2015petmri} and 2) directions~\cite{Gallardo2003, Gallardo2004, Haber2013, Ehrhardt2014tip, Ehrhardt2015petmri, Rigie2015tnvct, Ehrhardt2016mri, Ehrhardt2016pet, Knoll2016, Schramm2017petplusmri, Bathke2017mpi, Bungert2018remotesensing, Kolehmainen2019eit}. Real-world examples for these mathematical models are numerous as we will see in the next section.

\subsection{Application examples}
\begin{figure}%
\centering%
\includegraphics{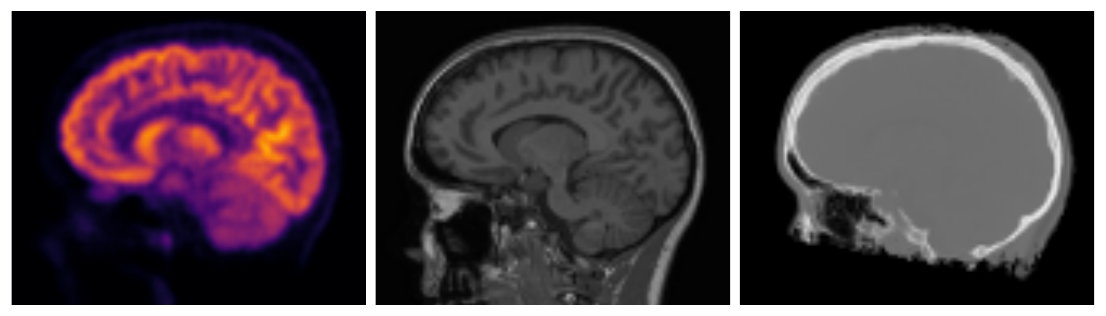}%
\caption{\textbf{PET-MR and PET-CT.} A low resolution functional PET image (left) is to be reconstructed with the help of an anatomical MRI (middle) or CT image (right). As is evident from the images, all three images share many edges due to the same underlying anatomy. Note that the high soft tissue contrast in MRI makes it favourable over CT for this application. Images curtesy of P. Markiewicz and J. Schott.} \label{fig:example:petmr}
\end{figure}

\begin{figure}%
\centering%
\includegraphics{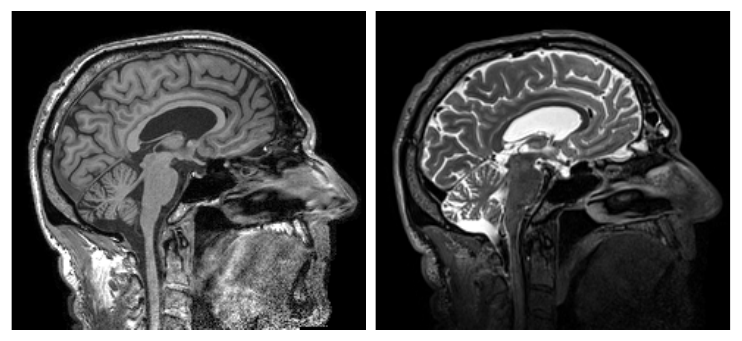}%
\caption{\textbf{Multi-contrast MRI.} The same MRI scanner can produce different images depending on the acquisition sequence such as $T_1$-weighted (left) and $T_2$-weighted images (right). Images courtesy of N. Burgos.} \label{fig:example:multicontrastmri}
\end{figure}

Historically the first application where information from several modalities was combined was positron emission tomography (PET) and magnetic resonance imaging (MRI) in the early 1990’s \cite{Leahy1991}. Sharing information between two different imaging modalities is motivated by the fact that all images will be highly influenced by the same underlying anatomy, see Figure \ref{fig:example:petmr}. Since single-photon emission computed tomography (SPECT) imaging is both mathematically and physically similar to PET imaging, most of the proposed models can be directly translated and often models are proposed for both modalities simultaneously, see e.g.~\cite{Bowsher1996, Rangarajan2000, Chan2007, Nuyts2007}. Over the years there always has been research in this direction, see e.g.~\cite{Bowsher1996, Rangarajan2000, Comtat2002, Bowsher2004, Baete2004, Chan2007, Chan2009, Tang2009, Bousse2010, Pedemonte2011a, Somayajula2011, Cheng-Liao2011, Vunckx2012, Kazantsev2012, Bousse2012, Bai2013}, which was intensified with the advent of the first simultaneous PET-MR scanner in 2011~\cite{Delso2011petmri}, see e.g. \cite{Knoll2014, Ehrhardt2014, Ehrhardt2015petmri, Tang2015, Ehrhardt2016pet, Knoll2016, Schramm2017petplusmri, Mehranian2017petmri, Mehranian2017petplusmri, Tsai2018mic, Zhang2018, Ehrhardt2019pmb, Deidda2019}.

The same motivation applies to other medical imaging techniques, for example multi-contrast MRI, see e.g.~\cite{Bilgic2011, Ehrhardt2016mri, Huang2014multicontrastmri, Sodickson2015, Song2018multicontrastmri, Xiang2019}. In multi-contrast MRI multiple acquisition sequences are used to acquire data of the same patient, see Figure~\ref{fig:example:multicontrastmri} for a $T_1$- and a $T_2$-weighted image with share anatomy. Other special cases are the combination of anatomical MRI (e.g. $T_1$-weighted) and magnetic particle imaging~\cite{Bathke2017mpi}, functional MRI (fMRI) and anatomical MRI~\cite{Rasch2018fmri} as well as anatomical (${}^{1}$H) and fluorinated gas (${}^{19}$F) MRI~\cite{Obert2020}. A related imaging task is quantitative MRI (such as Magnetic Resonance Fingerprinting~\cite{Ma2013})~\cite{Davies2013, Tang2018, Dong2019qmri, Golbabaee2020qmri} where one aims to reconstruct quantitative maps of tissue parameters (e.g. $T_1$, $T_2$, proton density, off-resonance frequency), but regularisers coupling these maps have not been used to date. The idea to couple channels has also been used for parallel MRI~\cite{Chen2013}.

\begin{figure}%
\centering%
\includegraphics{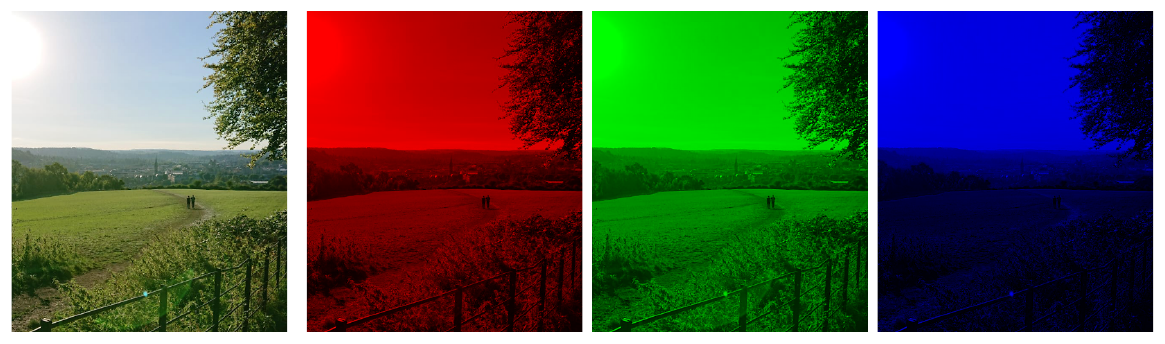}%
\caption{\textbf{Color imaging.} \index{Color imaging} The color image (left) is composed of three color channels (right) all of which show similar edges due to the same scenery. Images courtesy of M. Ehrhardt.} \label{fig:example:colorimage}
\end{figure}

\begin{figure}%
\centering%
\includegraphics{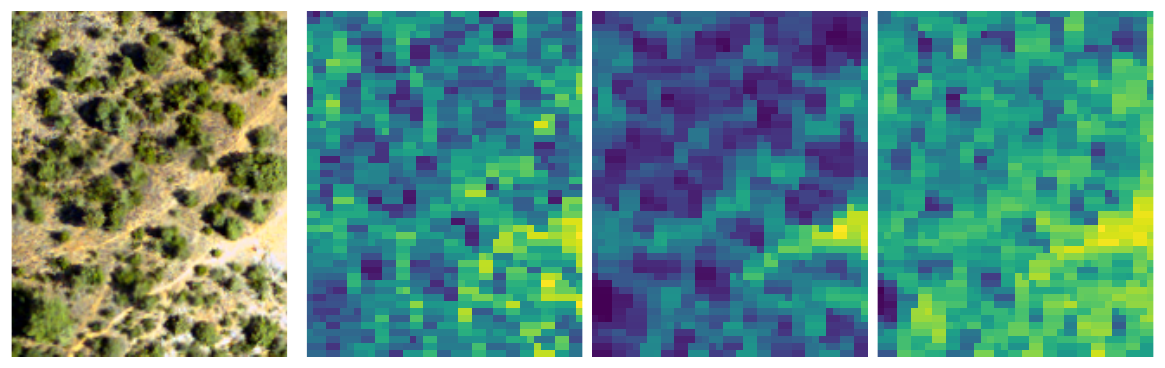}%
\caption{\textbf{Hyperspectral imaging + photography.} \index{Pansharpening}\index{Remote sensing} A nowadays common scenario is that multiple cameras are mounted on a plane or satellite for remote sensing. While one camera carries spectral information (right), the other has high spatial resolution (left). Images courtesy of D. Coomes.} \label{fig:example:remotesensing}
\end{figure}

\begin{figure}
\centering%
\includegraphics{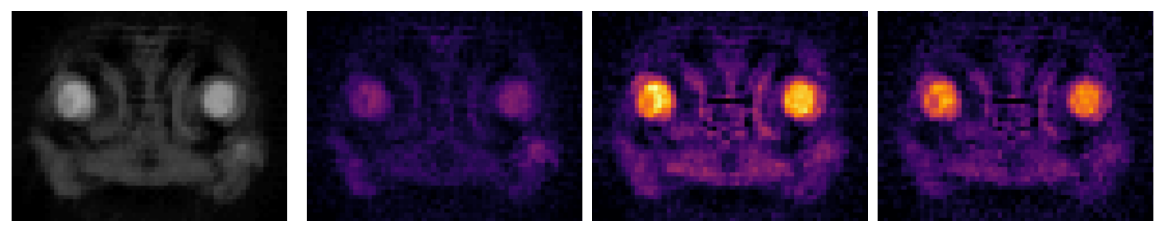}%
\caption{\textbf{Spectral CT.} \index{Spectral CT} Standard (white-beam) CT on the left and three channels (28, 34 and 39 keV) of spectral CT on the right of an iodine-stained lizard head reconstructed by CIL \cite{Ametova2019cil}. The spectral channels clearly show a large increase in intensity from 28 to 34 keV, thereby revealing the presence, location and concentration of iodine. Images courtesy of J. Jorgensen and R. Warr.} \label{fig:example:ct}
\end{figure}

Starting from the 1990’s, mathematical models were developed that make use of the expected correlations between colour channels of RGB images~\cite{Sapiro1996, Blomgren1998, Sochen1998}, see Figure \ref{fig:example:colorimage}. Research in this field is still very active today, see e.g.~\cite{Tschumperle2005, Bresson2008, Goldluecke2012, Holt2014tnv, Ehrhardt2014tip, Moeller2014}.

In remote sensing observations are often available from multiple sensors either mounted on a plane or on a satellite. For example a hyperspectral camera with low spatial resolution and a digital camera with higher spatial resolution may be used simultaneously, see Figure~\ref{fig:example:remotesensing}. This situation naturally invites for the fusion of information, see \cite{Ballester2006, Moeller2012, Fang2013, Loncan2015, Yokoya2017, Duran2017, Bungert2018remotesensing, Bungert2018robust} and references therein. In some situations the response of the cameras to certain wavelengths is (assumed to be) known such that the data can be fused making use of this knowledge. This is commonly referred to as, see e.g.  \textit{pansharpening} \cite{Loncan2015, Yokoya2017, Duran2017}. It is important to note that this assumption is sometimes not fulfilled and many of the aforementioned algorithms are flexible enough to fuse data in this more general situation.

Dual and spectral computed tomography (CT) is becoming increasingly popular in (bio-)~medical imaging and material sciences due to its ability to distinguish different materials which would not be possible using a single energy, see Figure~\ref{fig:example:ct}. Since the energy channels have a very different signal-to-noise ratio, coupling them within the reconstruction allows to transfer information from high signal to low signal channels~\cite{Rigie2015tnvct, FoygelBarber2016, Rigie2017, Kazantsev2018}.

In geophysics, the coupling between modalities has been used to model similarity between electrical resistivity and seismic velocity \cite{Gallardo2003, Gallardo2004}, estimating conductivity from multi-frequency data \cite{Haber1997}, inverting gravity and seismic tomography \cite{Haber1997} and controlled-source electromagnetic resistivity inversion~\cite{Meju2019}. For an overview and more details on examples in geophysics see in~\cite{Gallardo2011, Haber2013} and references therein.

Ideas from multi-modality imaging have recently also been used for art restoration. When a canvas is painted on both sides, an x-ray image shows the superposition of both paintings. The x-ray information can then be separated using photos of both sides of the canvas~\cite{Deligiannis2017art}.

Other examples that were considered in the literature are combining anatomical information and electrical impedance tomography~\cite{Kaipio1999, Kolehmainen2019eit}, CT and MRI~\cite{Xi2015}, photo-acoustic and optical coherence tomography \cite{Elbau2018}, x-ray fluorescence and transmission tomography \cite{Di2016} and various channels in multi-modal electron tomography~\cite{Huber2019}. The combination of various imaging modalities into one system may eventually lead to what is sometimes referred to as \textit{omni-tomography}~\cite{Wang2012}.

Image reconstruction with side information is mathematically similar to multi-modal image registration and thus it is not surprising that both fields share a lot of mathematical models, see e.g. \cite{Wells1996, Maes1997, Pluim2000, Haber2006}.

\subsection{Variational regularisation}
Inverse problems of the form \eqref{eq:ip} can be solved using variational regularisation, i.e. framed as the optimisation problem
\begin{equation}
u_\alpha \in \arg\min_u \mathcal D(A u, f) + \alpha \mathcal R(u) \, . \label{eq:varreg}
\end{equation}
Here the \textit{data fidelity} $\mathcal D : Y \times Y \to \mathbb R_\infty := \mathbb R \cup \{\infty\}$ measures how close the estimated data $A u$ fits the acquired data $f$. The \textit{regulariser} (also referred to as the \textit{prior}) $\mathcal R : X \rightarrow \mathbb R_\infty$ defines which properties of the image $u$ we favour and which we do not. The trade-off between data fitting and regularisation can be chosen using the \emph{regularisation parameter} $\alpha > 0$. Problems of this form have been extensively studied, see for instance \cite{Engl1996, Scherzer2008book, Ito2014book, Bredies2018book, Benning2018actanumerica} and references therein.

Three popular regularisers for imaging are the \emph{squared $H^1$-semi norm}~($\Hone$), the \emph{total variation}~($\TV$) \cite{Rudin1992ROF, Burger2013} and the \emph{total generalised variation}~($\TGV$)~\cite{Bredies2010, Bredies2014, Bredies2015tgvnumerics}. It is common to model images as functions $u : \Omega \subset \mathbb R^d \to \mathbb R$. If $u$ is smooth enough, then these regularisers are defined as
\begin{align}
\Hone(u) &= \int_\Omega |\nabla u(x)|^2 \; \mathrm{d}x \label{eq:h1} \\
\TV(u) &= \int_\Omega |\nabla u(x)| \; \mathrm{d}x \label{eq:tv} \\
\TGV (u) &= \inf_\zeta \int_\Omega |\nabla u(x) - \zeta(x)| + \beta |E \zeta(x)| \; \mathrm{d}x \label{eq:tgv} \, .
\end{align}
Here $\nabla u : \Omega \to \mathbb R^d, [\nabla u]_i = \partial_i u$ denotes the gradient of $u$, $E \zeta : \Omega \to \mathbb R^{d \times d}, [E \zeta]_{i,j} = (\partial_i \zeta_j + \partial_j \zeta_i)/2$ denotes the symmetrised gradient of a vector-field $\zeta : \Omega \to \mathbb R^d$, see \cite{Bredies2015tgvnumerics} for more details, and $|\cdot|$ denotes the Euclidean/Frobenius norm. For $\TV$ and $\TGV$ it is of interest to develop other formulations which are well-defined even when $u$ is not smooth. For simplicity, we do not go into more detail in this direction but refer the interested reader to the literature, e.g. \cite{Bredies2010, Burger2013}.

All three regularisers promote solutions with different smoothness properties. $\Hone$ promotes smooth solutions with small gradients everywhere, whereas $\TV$ promotes solutions which have sparse gradients, i.e. the images are piecewise constant and appear cartoon-like. The latter also leads to the staircase artefact which can be overcome by $\TGV$ which promotes piecewise linear solutions. None of these regularisers are able to encode additional information on the location or direction of edges.

\subsection{Contributions}

The contributions in this chapter are threefold.

\runinhead{Overview over existing methods} We provide an overview on existing mathematical models for structural similarity which are related to the shared location or direction of edges. We then discuss various regularisers which promote similarity in this sense.

\runinhead{Higher order models} Existing methods focus on incorporating additional information into regularisers modelling first-order smoothness. We extend existing methodology to second-order smoothness using the total generalised variation framework.

\runinhead{Extensive numerical comparison} We highlight the properties of the discussed regularisers and the dependence on various parameters using two inverse problems: tomography and super-resolution.

\subsection{Related work}

\subsubsection{Joint reconstruction}
One can think of the setting \eqref{eq:ip} with extra information $v$ as a special case when multiple measurements
\begin{equation}
K_i u_i = f_i \quad i = 1, \dots, m \label{eq:multiple:ip}
\end{equation}
are taken. If $m=2$ and one inverse problem is considerably less ill-posed, then this can be solved first to guide the inversion of the other. Some of the described models can be extended to the more general case (e.g. an arbitrary number of modalities) or the joint recovery of both/all unknowns, see e.g. \cite{Sapiro1996, Haber1997, Arridge1997probdiffusion, Gallardo2003, Gallardo2004, Gallardo2011, Chen2013, Haber2013, Knoll2014, Ehrhardt2014tip, Holt2014tnv, Ehrhardt2015petmri, Rigie2015tnvct, Knoll2016, Di2016, Mehranian2017petmri, Zhang2018, Meju2019, Huber2019}, but it is out of the scope of this chapter to provide an overview on those. For an overview up to 2015, see \cite{Ehrhardt2015phd}. A few recent contributions are summarized in \cite{Arridge2020preface}.

Model~\eqref{eq:multiple:ip} may include several special cases i) multiple measurements of the same unknown, i.e. $u_i = u$ and ii) measurements correspond to different states of the same unknown, e.g. in dynamic imaging $u_i = u(\cdot, t_i)$.
The former case is covered by the standard literature when concatenating the measurements and the systems models, i.e. $(Ku)_i := K_i u$ and $f = (f_1, \ldots, f_m)$. The latter has been widely studied in the literature, too, see e.g.~\cite{Schmitt2002a, Schmitt2002b, Schuster2018dynamicip} and references therein. Both of these are in general unrelated to multi-modality imaging.

\subsubsection{Other models for similarity}
The earliest contributions to structure-promoting regularisers for multi-modality imaging were made in the early 1990’s by Leahy and Yan \cite{Leahy1991} who used a segmentation of an anatomical MRI image to enhance PET reconstruction. This is achieved by carefully handcrafting a regulariser which can encode this information. In this chapter we will use the same strategy but in a continuous setting which is independent of the discretisation and will not rely on a segmentation of the side information $v$. These ideas were subsequently refined in various directions \cite{Bowsher1996, Rangarajan2000, Comtat2002, Bowsher2004, Baete2004, Chan2007, Chan2009, Bousse2010, Pedemonte2011a, Bilgic2011, Bousse2012, Bai2013} of which Bowsher’s prior \cite{Bowsher2004} remains most popular today.

Other models that can combine information of multiple modalities are based on coupled diffusion \cite{Arridge1997probdiffusion, Tschumperle2005, Arridge2008}, level-sets \cite{Cheng-Liao2011}, information theoretic priors (joint entropy, mutual information) \cite{Nuyts2007, Tang2009, Somayajula2011, Tang2015}, Bregman distances \cite{Ballester2006, Moeller2012, Estellers2013, Kazantsev2014, Rasch2018fmri}, Bregman iterations \cite{Moeller2014, Rasch2018petmr}, the structure tensor \cite{Estellers2015}, joint dictionary learning \cite{Deligiannis2017art, Song2018multicontrastmri, Song2019}, common edge weighting \cite{Zhang2018} and deep learning \cite{Xiang2019}. Most of these methods are very different to what will be described in this chapter. There are some similarities between the methods of this chapter and methods which are based on the Bregman distance of the total variation \cite{Ballester2006, Moeller2012, Estellers2013, Moeller2014, Kazantsev2014, Rasch2018petmr, Rasch2018fmri} but a detailed treatment is outside the scope of this section.

\section{Mathematical models for structural similarity} \label{section:similarity}
In this section we define mathematical models where we aim to capture the similarities as shown in Figures \ref{fig:example:petmr} to \ref{fig:example:ct}. We start by explicitly stating two definitions which capture structural similarity which have been used implicitly in the literature. The first is based on the location of edges or the edge set \cite{Arridge2008, Bresson2008, Haber2013, Chen2013, Knoll2014, Moeller2014, Ehrhardt2015petmri, Zhang2018} and the second is based on direction of edges or the shape of an object \cite{Gallardo2003, Gallardo2004, Haber2013, Ehrhardt2014tip, Ehrhardt2015petmri, Rigie2015tnvct, Knoll2016}. The latter is essentially the same as Definition 5.1.6 in \cite{Ehrhardt2015phd} except for the degenerate case when either $\nabla u(x) = 0$ or $\nabla v(x) = 0$.

\begin{definition}[Structural similarity with edge sets] \label{def:edgesets}%
Two differentiable images $u, v : \Omega \rightarrow \mathbb R$ are said to be \emph{structurally similar in the sense of edge sets} if
\begin{equation}
\mathcal Eu = \mathcal Ev
\end{equation}
where $\mathcal Eu = \{x \in \Omega \mid \nabla u(x) \neq 0\}.$ We also write $u \esim v$ to denote that $u$ and $v$ are structurally similar in the sense of edge sets.
\end{definition}

\begin{definition}[Structural similarity with parallel level sets] \label{def:pls}%
Two differentiable images $u, v : \Omega \rightarrow \mathbb R$ are said to be \emph{structurally similar in the sense of parallel level sets} if $u \esim v$ and for all $x \in \mathcal Eu$ there is
\begin{equation}
\nabla u(x) \parallel \nabla v(x) \, .
\end{equation}
We also write $u \dsim v$ to denote that $u$ and $v$ are structurally similar in the sense of parallel level sets.
\end{definition}

\begin{remark}
For smooth images $u$ and $v$, their gradients are perpendicular to their level sets, i.e. $u^{-1}(s) = \{ x \in \Omega \mid u(x) = s\}$. Thus parallel gradients is equivalent to parallel level sets which explains the naming. The notion that the structure of an image is contained in its level sets dates back to \cite{Caselles2002}.
\end{remark}

\begin{remark} 
By definition, similarity with parallel level sets (Definition \ref{def:pls}) is stronger than the definition that only involves edge sets (Definition \ref{def:edgesets}). An example of two images $u$ and $v$ which have the same edge set but do not have parallel level sets is the following. $u, v : \Omega \subset \mathbb R^2 \to \mathbb R, u(x) = x_1, v(x) = x_2$. Clearly they have the same edge set since $\mathcal Eu = \mathcal Ev = \Omega$, but they do not have parallel level sets since $\nabla u(x) = [1, 0]$ but $\nabla v(x) = [0, 1]$.
\end{remark}

\begin{remark} Two images $u$ and $v$ have parallel level sets if and only if $u \esim v$ and for all $x \in \mathcal Eu$ there exists $\alpha \in \mathbb R$ such that
\begin{equation}
\nabla u(x) = \alpha \nabla v(x) \, .
\end{equation}
Examples of images which have parallel level sets include:
\begin{enumerate}
\item \emph{Function value transformations}. Let $f : \mathbb R \rightarrow \mathbb R$ be smooth and strictly monotonic, i.e. $f^\prime > 0$ or $f^\prime < 0$. Then $v := f \circ u \dsim u$. This is readily to be seen from the fact that $\nabla v(x) = f^\prime(u(x)) \nabla u(x) \neq 0$ if and only if $\nabla u(x) \neq 0$.

\item \emph{Local function value transformations}. Let $f_i : \mathbb R \rightarrow \mathbb R$ be smooth and strictly monotonic and $u = \sum_i u_i$ where $u_i$ are smooth functions whose gradients have mutually disjoint support. Then $v := \sum_i f_i \circ u_i \dsim u$.
\end{enumerate}
\end{remark}

\begin{remark}
It has been argued in the literature that many multi-modality images $z : \Omega \to \mathbb R^m$ essentially decompose as
\begin{align}
z_i(x) = \tau_i(x) \rho(x)
\end{align}
where $\rho(x)$ describes its structure and $\tau$ is a material property, see e.g. \cite{Kimmel2000, Holt2014tnv}. Since the material does not change arbitrarily, it is natural to assume that $\tau_i$ is slowly varying or even piecewise constant. In the latter case, if $x$ is such that $\nabla \tau_i(x) = 0$, then we have
\begin{align}
\nabla z_i(x) = \tau_i(x) \nabla \rho(x) \, , 
\end{align}
in particular if $\tau_i, \tau_j \neq 0$, then $z_i \dsim z_j$. This property is also related to the material decomposition in spectral CT, see e.g. \cite{Fessler2002, Heismann2012spectralCT, Long2014}.
\end{remark}

\subsection{Measuring structural similarity}
Measuring the degree of similarity with respect to the previous two definitions of structural similarity is not easy and we will now discuss a couple of ideas from the literature. Here and for the rest of this chapter, we will make frequent use of the vector-valued representation of a set of images $z : \Omega \to \mathbb R^2, z(x) := [u(x), v(x)]$. We denote by $J$ its Jacobian, i.e. $J : \Omega \to \mathbb R^{d \times 2}, J_{i, j} = \partial_i z_j$.

With the definition of the Jacobian we see that $u \esim v$ if and only if
\begin{align}
\int_\Omega |J(x)|_0 \; \mathrm{d}x = \int_\Omega |\nabla u(x)|_0 \; \mathrm{d}x = \int_\Omega |\nabla v(x)|_0 \; \mathrm{d}x \label{eq:Jacobian:zero}
\end{align}
where $|x|_0 := 1$ if $x \neq 0$ and $0$ else.

Similarly, by definition $u \dsim v$ if and only if $u \esim v$ and (a) $\operatorname{rank} J(x) = 1$ for all $x \in \mathcal Eu$. (a) is equivalent to (b) a vanishing determinant, i.e. $\det J^\top\!(x)J(x) = 0$. Simple calculations, see e.g. \cite{Ehrhardt2015phd}, show that 
\begin{align}
\det J^\top\!(x)J(x) = |\nabla u(x)|^2 |\nabla v(x)|^2 - \langle \nabla u(x), \nabla v(x) \rangle^2 \, , \label{eq:det}
\end{align}
where we use the notation $\langle x, y \rangle = x^\top y$ for the inner product of two column vectors $x$ and $y$. In order to get further equivalent statements we turn to the singular values of the Jacobian which are given by 
\begin{align}
\sigma_{1/2}^2(x) = \frac12 \left[|J(x)|^2 \pm \sqrt{|J(x)|^4 - \det J^\top\!(x)J(x)} \right]
\end{align}
with $|J(x)|^2 = |\nabla u(x)|^2 + |\nabla v(x)|^2$, see e.g. \cite{Ehrhardt2015phd}. Since $\sigma_1(x) \geq \sigma_2(x) \geq 0$ we have that (a) holds if and only if (c) the second singular value vanishes, i.e. $\sigma_2(x) = 0$ or (d) the vector of singular vectors $\sigma(x) = [\sigma_1(x), \sigma_2(x)]$ is 1-sparse.

\section{Structure-promoting regularisers} \label{sec:regularizers}

Many of the abstract models from the previous section to measure the degree of similarity with respect to the previous two definitions of structural similarity are computationally challenging as they relate to non-convex constraints. In this section we will define convex structure-promoting regularisers which make them computationally tractable.

\subsection{Isotropic models}
We first look at isotropic models which only depend on gradient magnitudes rather than directions, thus promote structural similarity in the sense of edge sets, Definition~\ref{def:edgesets}.

\begin{figure}%
\centering%
\includegraphics{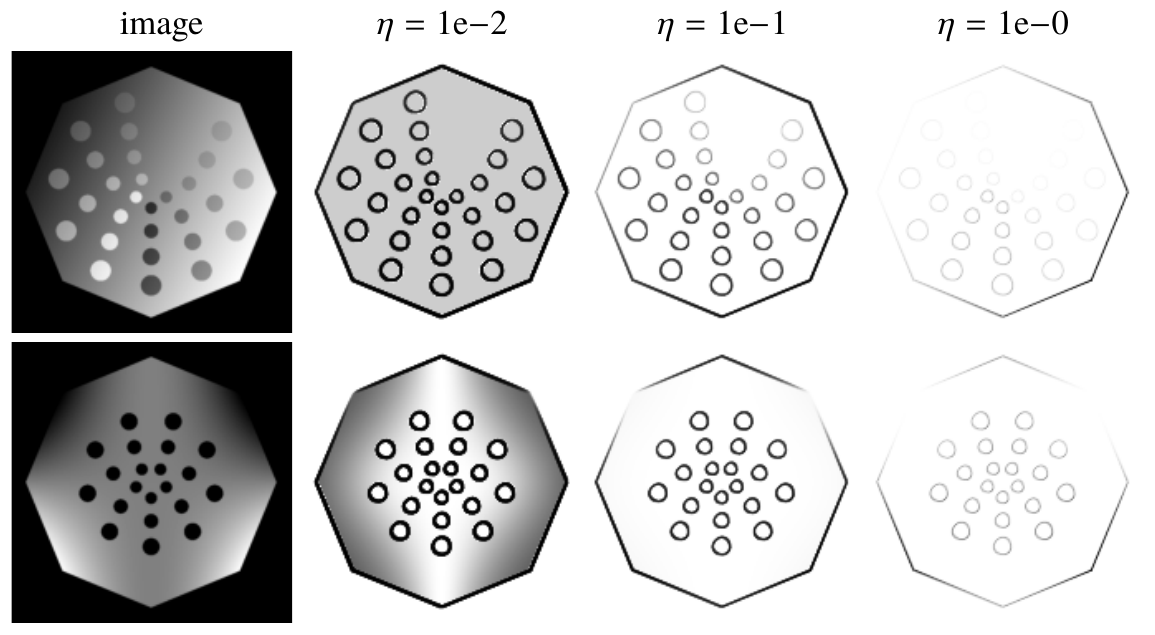}%
\caption{Influence of the parameter $\eta$ on estimation of \textbf{edge location}. The images on the right show the scalar field $w : \Omega \to [0, 1]$ which locally weights the influence of the regulariser, see \eqref{eq:edge:w}. Here "black" denotes 0 and "white" denotes 1.} \label{fig:example:location}
\end{figure}

First, based on \eqref{eq:Jacobian:zero} if we approximate $|J(x)|_0$ by $|J(x)|$, then 
\begin{align}
\JTV(u) &= \int_\Omega |J(x)| \; \mathrm{d}x = \int_\Omega \sqrt{|\nabla u(x)|^2 + |\nabla v(x)|^2} \; \mathrm{d}x \\
&\leq \int_\Omega |\nabla u(x)| + |\nabla v(x)| \; \mathrm{d}x
= \TV(u) + \TV(v) \label{eq:jtv}
\end{align}
with equality if and only if $\mathcal Eu \cap \mathcal Ev = \emptyset$. This regulariser is called \emph{joint total variation} in some communities, see e.g. \cite{Chen2013, Haber2013, Ehrhardt2015petmri, Ehrhardt2016pet} and \emph{vectorial total variation} in others, see e.g. \cite{Bresson2008}.

\begin{remark}
Note that $\JTV$ has the favourable property that if $\nabla v = 0$, then
$\JTV(u) = \TV(u)$, so that it reduces to a well defined regularisation in $u$ in this degenerate case. Note that this property also holds locally.
\end{remark}

\begin{remark}
We would also like to note that there is a connection between $\JTV$ and the singular values of $J$. Let $\sigma_1, \sigma_2 : \Omega \to [0, \infty)$ be the two singular values of $J$, then we have
\begin{align}
\JTV(u) &= \int_\Omega \sqrt{\sigma_1^2(x) + \sigma_2^2(x)} \; \mathrm{d}x \, .
\end{align}
\end{remark}

Another strategy to favour edges at similar locations while reducing to a well-defined regulariser in the degenerate case is to introduce local weighting. Let $w : \Omega \to [0, 1]$ be an edge indicator function for $v$ such that $w(x) = 1$ when $\nabla v(x) = 0$ and a small value whenever $|\nabla v(x)|$ is large. For example, choose
\begin{align}
w(x) = \frac{\eta}{\sqrt{\eta^2 + |\nabla v(x)|^2}} \label{eq:edge:w}
\end{align}
which is illustrated in Figure~\ref{fig:example:location}. The figure shows that with a medium $\eta$ the weight $w$ in \eqref{eq:edge:w} shows the main structures of the images so that these can be promoted in the other image. If $\eta$ is too small, then also unwanted structures are captured in $w$ such as a smooth background variation. If $\eta$ is too large, then the structures start to disappear.

For regularisers which are based on the image gradient $\nabla u$, the weighting $w$ can be used to favour edges at certain locations by replacing $\nabla$ by $w \nabla$. For instance, for $\Hone$ \eqref{eq:h1}, $\TV$ \eqref{eq:tv} and $\TGV$ \eqref{eq:tgv} this strategy results in 
\begin{align}
\wHone(u) &= \int_\Omega |w(x) \nabla u(x)|^2 \; \mathrm{d}x = \int_\Omega w^2(x) |\nabla u(x)|^2 \; \mathrm{d}x \label{eq:wh1} \\
\wTV(u) &= \int_\Omega |w(x) \nabla u(x)| \; \mathrm{d}x = \int_\Omega w(x) |\nabla u(x)| \; \mathrm{d}x \label{eq:wtv} \\
\wTGV(u) &= \inf_\zeta \int_\Omega |w(x) \nabla u(x) - \zeta(x)| + \beta |E \zeta(x)| \; \mathrm{d}x \label{eq:wtgv}
\end{align}
which we will refer to as \emph{weighted squared $H^1$-semi norm}, \emph{weighted total variation} and \emph{weighted total generalised variation}. $\wTV$ was used in \cite{Arridge2008, Lenzen2015, Ehrhardt2016mri}. A variant of $\wTV$ has been considered for single modality imaging in \cite{Hintermuller2010, Dong2011} and extended to a variant of $\wTGV$ \cite{Bredies2012}.

\begin{remark} 
The parameter $\eta$ in $w$, see \eqref{eq:edge:w}, should be chosen in relation to $|\nabla v(x)|$. A common strategy is to normalise the side information first such that $\sup_{x \in \Omega} |\nabla v(x)| = 1$. Then desirable values of $\eta$ are usually within the range $[0.01, 1]$.
\end{remark}

\subsection{Anisotropic models}

The same idea which resulted in isotropically "weighted" variants of common regularisers can be used anisotropically, i.e. by making the local weights vary with direction. Let us denote the anisotropic weighting by $D : \Omega \to \mathbb R^{d \times d}$. Similar to the isotropic variant, one would like the weight to become the identity matrix, i.e. $D(x) = I$, when $\nabla v(x) = 0$. In order to promote parallel level sets it is desirable that $D(x)\nabla u(x)$ should be small if $\nabla u(x) \parallel \nabla v(x)$ and $D(x)\nabla u(x) = \nabla u(x)$ if $\nabla u(x) \perp \nabla v(x)$. For example
\begin{align}
D(x) = I - \gamma \xi(x)\xi^\top\!(x) \, , \quad \xi(x) = \frac{\nabla v(x)}{\sqrt{\eta^2 + |\nabla v(x)|^2}}\label{eq:edge:d}
\end{align}
for $\gamma \in (0, 1]$ (usually close to 1) and $\eta > 0$ satisfies all of these properties. Clearly if $\nabla v(x) = 0$ then $\xi = 0$ such that $D(x) = I$. Moreover, if $\nabla u(x) \parallel \nabla v(x)$, then there exists an $\alpha$ such that $\nabla u(x) = \alpha \nabla v(x)$ and
\begin{align}
D(x)\nabla u(x) 
&= \left[I - \frac{\gamma}{\eta^2 + |\nabla v(x)|^2} \nabla v(x) \nabla v^\top\!(x)\right] \nabla u(x) \\
&= \left[1 - \frac{\gamma |\nabla v(x)|^2}{\eta^2 + |\nabla v(x)|^2}\right] \nabla u(x) \, .
\end{align}
The scalar weighting factor converges to $1 - \gamma$ for $|\nabla v(x)| \rightarrow \infty$. Finally, if $\nabla u(x) \parallel \nabla v(x)$, then clearly $D(x)\nabla u(x) = \nabla u(x)$.

\begin{figure}
\centering%
\includegraphics{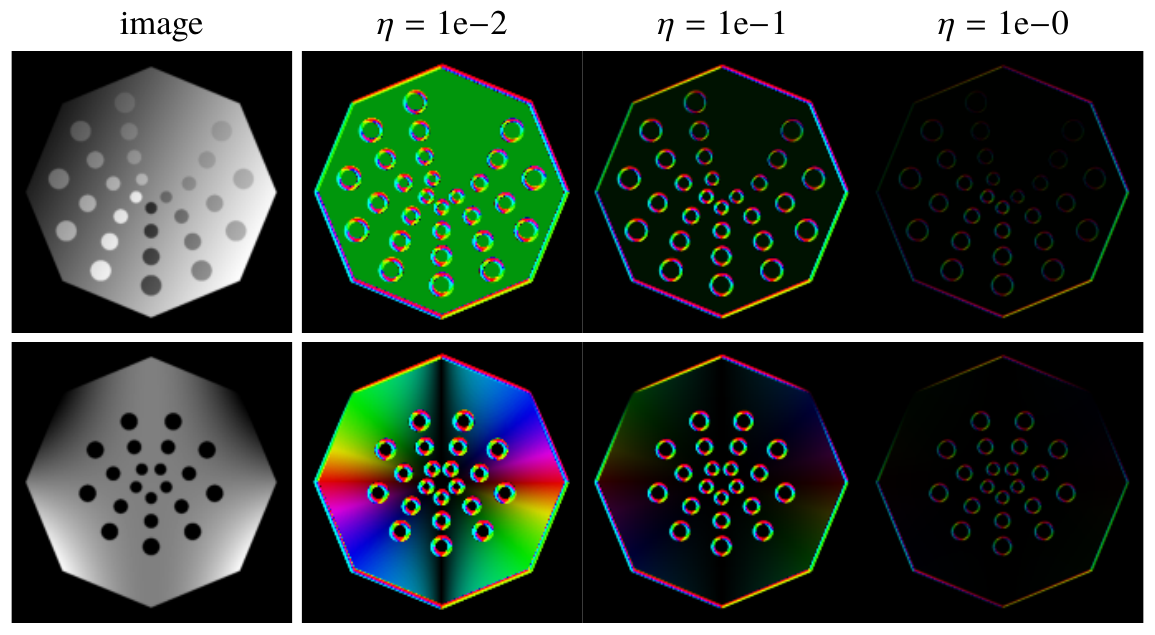}%
\caption{Influence of the parameter $\eta$ on estimation of \textbf{edge location and direction}. The images on the right show the vector field $\xi : \Omega \to \mathbb R^d$ which locally defines the influence of the regulariser, see e.g. \eqref{eq:edge:d}. Here "black" denotes that the magnitude of $\xi$, i.e. $|\xi(x)|$, is 0 and a bright colour denotes that $|\xi(x)|$ is 1. The colours show the direction of the vector field $\xi$ modulo its sign.} \label{fig:example:direction}
\end{figure}

The example of the matrix-field $D : \Omega \to \mathbb R^{d \times d}$ in \eqref{eq:edge:d} is determined by the vector-field $\xi : \Omega \to \mathbb R^d$ which we visualise in Figure \ref{fig:example:direction}. The colours show the direction of the vector-field modulo its sign (since $\xi(x)\xi^\top\!(x)$ is invariant to a change of sign) and the brightness indicate its magnitude $|\xi(x)|$. Note that images appear as colour versions of Figure \ref{fig:example:location} which shows the isotropic weighting $w$.

Using a matrix-field in common regularisers lead to their "directional" variant
\begin{align}
\dHone(u) &= \int_\Omega | D(x) \nabla u(x)|^2 \; \mathrm{d}x \label{eq:dh1} \\
\dTV(u) &= \int_\Omega | D(x) \nabla u(x)| \; \mathrm{d}x \label{eq:dtv} \\
\dTGV(u) &= \inf_\zeta \int_\Omega |D(x) \nabla u(x) - \zeta(x)| + \beta |E \zeta(x)| \; \mathrm{d}x \label{eq:dtgv} \, .
\end{align}

\begin{remark}
There is a connection between the particular choice of the matrix-field $D$ in \eqref{eq:edge:d} and the Jacobian $J$.
\begin{align}
|D(x)\nabla u(x)|^2
&= |\nabla u(x) - \frac{\gamma}{\eta^2 + |\nabla v(x)|^2} \langle \nabla u(x), \nabla v(x)\rangle \nabla v(x)|^2 \\
&= |\nabla u(x)|^2 - \frac{2 \gamma \eta^2 + \gamma (2 - \gamma) |\nabla v(x)|^2}{(\eta^2 + |\nabla v(x)|^2)^2} \langle \nabla u(x), \nabla v(x)\rangle^2 \, .
\end{align}
For $\eta = 0$, $\gamma = 1$ and $|\nabla v(x)| = 1$, then with \eqref{eq:det} we have
\begin{align}
|D(x)\nabla u(x)|^2 = |\nabla u(x)|^2 |\nabla v(x)|^2 - \langle \nabla u(x), \nabla v(x)\rangle^2 = \det J^\top\!(x)J(x) \, .
\end{align}
Thus, $\dHone$ corresponds to penalising the determinant. This regulariser is widely used for joint reconstruction in geophysics under the name \emph{cross-gradient} function since it is also the cross product of $\nabla u(x)$ and $\nabla v(x)$, see e.g. \cite{Gallardo2003, Gallardo2004, Gallardo2011, Meju2019}.
Similarly the $\dTV$ used for instance in medical imaging \cite{Ehrhardt2016mri, Ehrhardt2016pet, Bathke2017mpi, Schramm2017petplusmri, Kolehmainen2019eit, Obert2020} and remote sensing \cite{Bungert2018remotesensing} can be seen as penalising the square root of the determinant.
\end{remark}

Another strategy to promote parallel level sets is via nuclear norm of the Jacobian which is defined as $|J(x)|_\ast = \sum_{i=1}^{\min(d, 2)} \sigma_i(x)$ where $\sigma_i(x)$ denotes the $i$th singular value of $J(x)$. Using the nuclear norm promotes sparse vectors of singular values $\sigma(x) = [\sigma_1(x), \sigma_2(x)]$ and thereby parallel level sets. As a regulariser 
\begin{align}
\TNV(u) = \int_\Omega |J(x)|_\ast \; \mathrm{d}x \label{eq:tnv}
\end{align}
this strategy became known as \emph{total nuclear variation}, see \cite{Holt2014tnv, Rigie2015tnvct, Knoll2016, Rigie2017}.

All first-order regularisers of this section can be readily summarised in the following standard form
\begin{equation}
\mathcal J(u) = \int_\Omega \phi[B(x) \nabla u(x)] \; \mathrm{d}x \label{eq:general}
\end{equation}
where $B(x) : \mathbb R^d \rightarrow \mathbb R^m$ is a an affine transformation and $\phi: \mathbb R^m \rightarrow \mathbb R$. For details how $B$ and $\phi$ can be chosen for specific regularisers to fit this framework, see Table \ref{tab:reg}. It is useful for Jacobian-based regularisers to use the reweighted Jacobian $[\nabla u(x), \xi(x)]$ with $\xi(x) = \eta \nabla v(x)$ instead.

\begin{table}
\centering
\caption{Examples of first-order structure-promoting regularisers, see \eqref{eq:general}.} \label{tab:reg}
\begin{tabular}{p{2.2cm}p{1.9cm}p{3.1cm}p{2cm}p{2cm}}
\hline\noalign{\smallskip}
regulariser & definition & $B(x)y$ & $m$ & $\phi(x)$  \\
\noalign{\smallskip}\hline\noalign{\smallskip}
$\Hone$ & \eqref{eq:h1} & $y$  & $d$ & $|x|^2$ \\
$\wHone$ & \eqref{eq:wh1} & $w(x) y$  & $d$ & $|x|^2$ \\
$\dHone$ & \eqref{eq:dh1} & $D(x) y$  & $d$ & $|x|^2$ \\ \noalign{\smallskip}\hline\noalign{\smallskip} 
$\TV$ & \eqref{eq:tv} & $y$  & $d$ & $|x|$ \\
$\wTV$ & \eqref{eq:wtv} & $w(x) y$  & $d$ & $|x|$ \\
$\dTV$ & \eqref{eq:dtv} & $D(x) y$  & $d$ & $|x|$ \\
$\JTV$ & \eqref{eq:jtv} & $[y, \xi(x)]$  & $d \times 2$ & $|x|$ \\
$\TNV$ & \eqref{eq:tnv} & $[y, \xi(x)]$  & $d \times 2$ & $|x|_\ast$ \\
\noalign{\smallskip}\hline\noalign{\smallskip}
\end{tabular}
\vspace*{-12pt}
\end{table}

\section{Algorithmic solution}
Note that the solution to variational regularisation \eqref{eq:varreg} with either first- \eqref{eq:general} or second-order structural regularisation \eqref{eq:tgv}, \eqref{eq:wtgv}, \eqref{eq:dtgv} can be cast into the general non-smooth composite optimisation form
\begin{equation}
\min_x \mathcal F(A x) + \mathcal G(x) \label{eq:genericoptimization}
\end{equation}
with $\mathcal F(y) = \sum_{i=1}^n \mathcal F_i(y_i)$ and $Ax = [A_1x, \ldots, A_nx]$, see Table \ref{tab:alg}. We denote by $\|\cdot\|_{2, 1}, \|\cdot\|_2^2$ and $\|\cdot\|_{*, 1}$ discretisations of 
\begin{align}
z \mapsto \int_\Omega |z(x)| \; \mathrm{d}x\; , \quad  z \mapsto \int_\Omega |z(x)|^2 \; \mathrm{d}x \quad  \text{and} \quad z \mapsto \int_\Omega |z(x)|_\ast \; \mathrm{d}x \, .
\end{align}

\subsection{Algorithm}
A popular algorithm to solve \eqref{eq:genericoptimization} and therefore \eqref{eq:varreg} is the primal-dual hybrid gradient (PDHG) \cite{Esser2010, Chambolle2011}, see Algorithm \ref{alg:pdhg}. It consists of two simple steps only involving basic linear algebra and the evaluation of the operator $A$ and its adjoint $A^*$. Moreover, it involves the computation of the proximal operator of $\tau \mathcal G$ and the convex conjugate of $\sigma \mathcal F^*$ where $\tau$ and $\sigma$ are scalar step sizes. The proximal operator of a functional $\mathcal H$ is defined as
\begin{equation}
\operatorname{prox}_{\mathcal H}(z) := \arg\min_x \left\{ \frac12 \|x - z\|_2^2 + \mathcal H(x) \right\} \label{eq:proximaloperator} \, .
\end{equation}

The proximal operator can be computed in closed-form for $\|\cdot\|_{2, 1}$ and $\|\cdot\|_2^2$. It also also be computed in closed-form for $\|\cdot\|_{*, 1}$ if either the number channels or the dimension of the domain are strictly less than 5, i.e. $m, d < 5$, see \cite{Holt2014tnv} for more details. Note also that the proximal operator of $\alpha \mathcal F(\cdot - \xi)$ can be readily computed based on the proximal operator of $\mathcal F$. More details on proximal operators, convex conjugates and examples can be found for example in \cite{Bauschke2011, Combettes2011prox, Parikh2014, Chambolle2016actanumerica}.

For some applications (e.g. x-ray tomography) a preconditioned \cite{Pock2011, Ehrhardt2019pmb} or randomised \cite{Chambolle2017, Ehrhardt2019pmb} variant can be useful but we will not consider these here for simplicity.

\begin{table}
\caption{Mapping the variational regularisation models into the composite optimisation framework \eqref{eq:genericoptimization}. In all cases we choose $A_1x = Ku$, $\mathcal F_1(y_1) = \mathcal D(y_1, b)$ and $\mathcal G(x) = \imath_{\geq 0}(u)$.}\label{tab:alg}
\centering
\begin{tabular}{p{1.6cm}p{1.5cm}p{1.2cm}p{1.7cm}p{1.2cm}p{2.28cm}p{1.47cm}}
\hline\noalign{\smallskip}
regulariser & definition & $x$ & $A_2x$ & $A_3x$ & $\mathcal F_2(y_2)$ & $\mathcal F_3(y_3)$  \\
\noalign{\smallskip}\svhline\noalign{\smallskip}
$\Hone$ & \eqref{eq:h1} & $u$ & $\nabla u$ & - & $\alpha \|y_2\|_2^2$ & - \\
$\wHone$ & \eqref{eq:wh1} & $u$ & $w\nabla u$ & - & $\alpha \|y_2\|_2^2$ & - \\
$\dHone$ & \eqref{eq:wh1} & $u$ & $D\nabla u$ & - & $\alpha \|y_2\|_2^2$& - \\ \noalign{\smallskip}\hline\noalign{\smallskip}
$\TV$ & \eqref{eq:tv} & $u$ & $\nabla u$ & - & $\alpha \|y_2\|_{2,1}$& - \\
$\wTV$ & \eqref{eq:wtv} & $u$ & $w \nabla u$ & - & $\alpha \|y_2\|_{2,1}$ & - \\
$\dTV$ & \eqref{eq:dtv} & $u$ & $D \nabla u$ & - & $\alpha \|y_2\|_{2,1}$ & - \\ \noalign{\smallskip}\hline\noalign{\smallskip}
$\JTV$ & \eqref{eq:jtv} & $u$ & $[\nabla u, 0]$ & - & $\alpha \|y_2 - [0, \xi]\|_{2, 1}$ & - \\
$\TNV$ & \eqref{eq:tnv} & $u$ & $[\nabla u, 0]$ & - & $\alpha \|y_2 - [0, \xi]\|_{\ast, 1}$ & - \\ \noalign{\smallskip}\hline\noalign{\smallskip}
$\TGV$ & \eqref{eq:tgv} & $(u, \zeta)$ & $\nabla u - \zeta$ & $E\zeta$ & $\alpha \|y_2\|_{2, 1}$ & $\alpha \beta \|y_3\|_{2, 1}$ \\
$\wTGV$ & \eqref{eq:wtgv} & $(u, \zeta)$ & $w\nabla u - \zeta$ & $E\zeta$ & $\alpha\|y_2\|_{2, 1}$ & $\alpha\beta \|y_3\|_{2, 1}$ \\
$\dTGV$ & \eqref{eq:dtgv} & $(u, \zeta)$ & $D\nabla u - \zeta$ & $E\zeta$ & $\mathcal \alpha \|y_2\|_{2, 1}$ & $\alpha \beta \|y_3\|_{2, 1}$ \\
\noalign{\smallskip}\hline\noalign{\smallskip}
\end{tabular}
\vspace*{-12pt}
\end{table}

\begin{algorithm}
\caption{Primal-dual hybrid gradient~(PDHG) to solve~\eqref{eq:genericoptimization}. Default values given in brackets.} \label{alg:pdhg}
\textbf{Input:} iterates $x (= 0)$, $y (= 0)$, step size parameter $\rho (=1)$ \\
\textbf{Initialize:} extrapolation $\overline{x} = x$, step sizes $\sigma = \rho / \|A\|$, $\tau = 0.999 / (\rho \|A\|)$
\begin{algorithmic}[1]
    \For{$k = 1, \ldots$}
    \State $x^+ = \operatorname{prox}_{\tau \mathcal G}\left(x - \tau A^*y\right)$
    \State $y^+ = \operatorname{prox}_{\sigma \mathcal F^\ast}\left(y + \sigma A (2x^+ - x)\right)$
    \EndFor
\end{algorithmic}%
\end{algorithm}

\subsection{Prewhitening}
Since the operator norms $\|A_i\|, i=1, \ldots n$ can vary significantly, it is often advisably to "prewhiten" the problem by recasting it as
\begin{equation}
\min_x \tilde {\mathcal F}(\tilde A x) + \mathcal G(x) \, .
\end{equation}
with $\tilde{\mathcal F}(y) := \sum_{i=1}^n \mathcal F_i(\|A_i\| \cdot y_i)$ and $\tilde A_i x := A_i x / \|A_i\|$. Then trivially $\|\tilde A_i\| = 1, i = 1, \ldots, n$ so that all operator norms are equal. Note that the proximal operator of $\sigma \tilde{\mathcal F}$ is simple to compute if the proximal operators of $\sigma \mathcal F_i, i = 1, \ldots, n$ are simple to compute, since
\begin{equation}
[\operatorname{prox}_{\sigma \tilde {\mathcal F}}(y)]_i = \lambda_i^{-1} [\operatorname{prox}_{\sigma \lambda_i^2 \mathcal F_i}(\lambda_i y_i)] \, ,
\end{equation}
for any $\lambda_i > 0$, see for instance \cite[Lemma 6.136]{Bredies2018book}.

\section{Numerical comparison}
This section describes numerical experiments to compare first- and second-order structure-promoting regularisers.

\subsection{Software, data and parameters}

\runinhead{Software} The numerical computations are carried out in Python using ODL (version 1.0.0.dev0) \cite{Adler2017odl} and ASTRA \cite{VanAarle2015, VanAarle2016} for computing line integrals in the tomography example. The source code which reproduces all experiments in this chapter can be found at \url{https://github.com/mehrhardt/Multi-Modality-Imaging-with-Structural-Priors}.

\begin{figure}
\centering%
\includegraphics{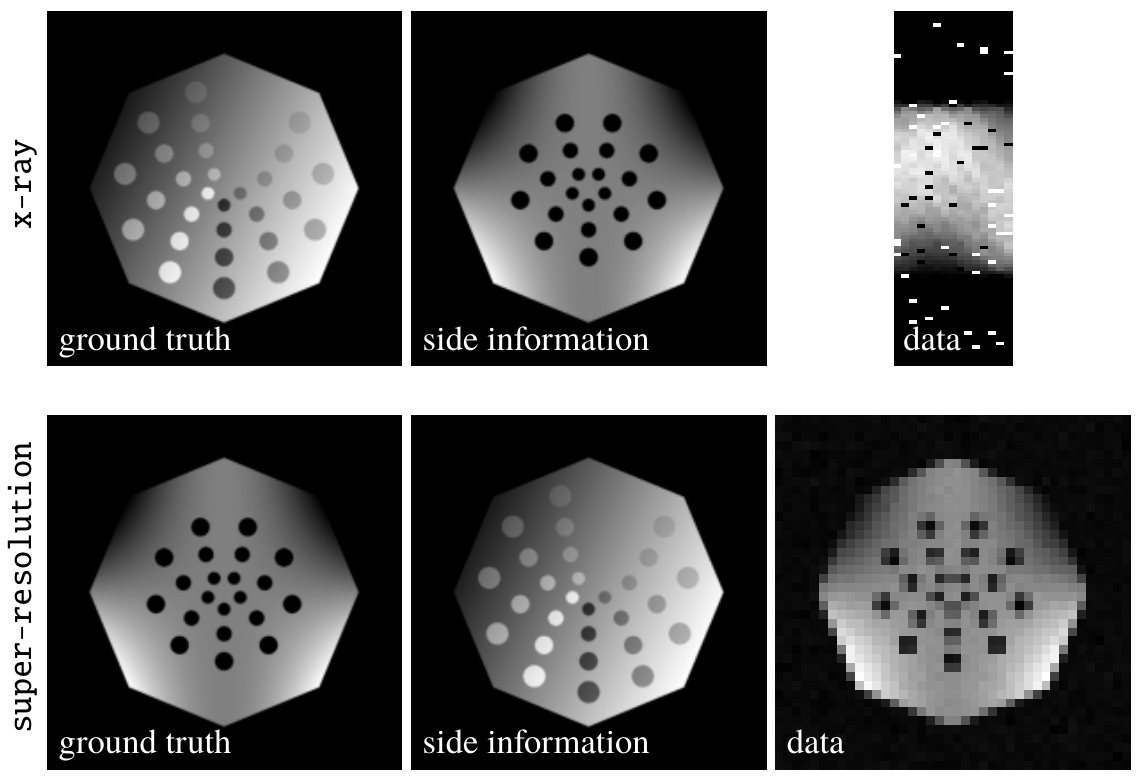}
\caption{\textbf{Test cases for numerical experiments}. Top: x-ray reconstruction from sparse views and failed detectors, bottom: super-resolution by a factor of 5 and Gaussian noise.} \label{fig:data}
\end{figure}

\runinhead{Data} We consider two test cases with different characteristics, both of which are visualised in Figure \ref{fig:data}. The first test case, later referred to as \dataxray, is parallel beam x-ray reconstruction from only 15 views where additionally some detectors are broken. The latter is modelled by salt-and-pepper noise where 5 \% of all detectors are corrupted. We aim to recover an image with domain $[-1, 1]^2$ discretised with $200^2$ pixels. The simulated x-ray camera has 100 detectors and a width of 3 in the same dimensions as the image domain. Therefore, the challenges are 1) sparse views, 2) small number of detectors and 3) broken detectors.

The second test case, which we refer to as \datasuper, considers the task of super-resolution. Also here we aim to recover an image with domain $[-1, 1]^2$ discretised with $200^2$ pixels. The forward operator is integrating over $5^2$ pixels, thus mapping images of size $200^2$ to images of size $40^2$. In addition, Gaussian noise of mean zero and standard deviation of 0.01 is added.

\runinhead{Algorithmic parameters} We chose the default value $\rho = 1$ for balancing the step sizes in PDHG and ran the algorithm for 3,000 iterations without choosing a specific stopping criterion.

\subsection{Numerical results}
The multiplicative scaling of an unconstrained optimisation problem is arbitrary, nevertheless we report the absolute values here for completeness. For simplicity, all regularisation parameters are shown as multiples of $\expnumber{1}{-4}$. The figures at the bottom right of each image are PSNR and SSIM.

\subsubsection{Test case \dataxray}

\runinhead{Effect of edge weighting}

\begin{figure}
\centering%
\includegraphics{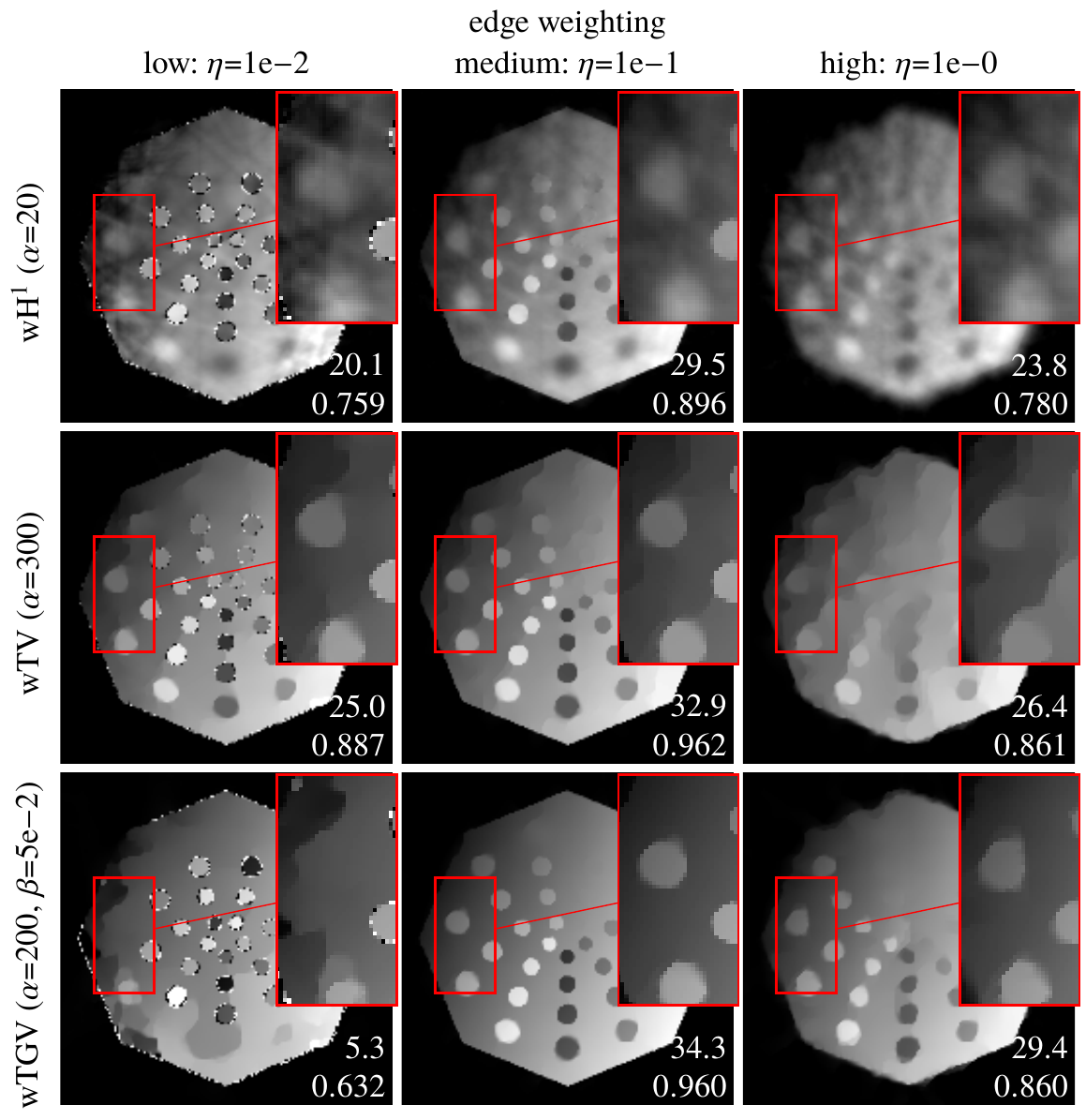}
\caption{Effect of edge weighting on locally weighted models for test case \dataxray: increasing edge parameter $\eta$ from left to right. All other parameters where tuned to maximize the PSNR and visual image quality.} \label{fig:xray:edge:w}
\end{figure}

\begin{figure}
\centering%
\includegraphics{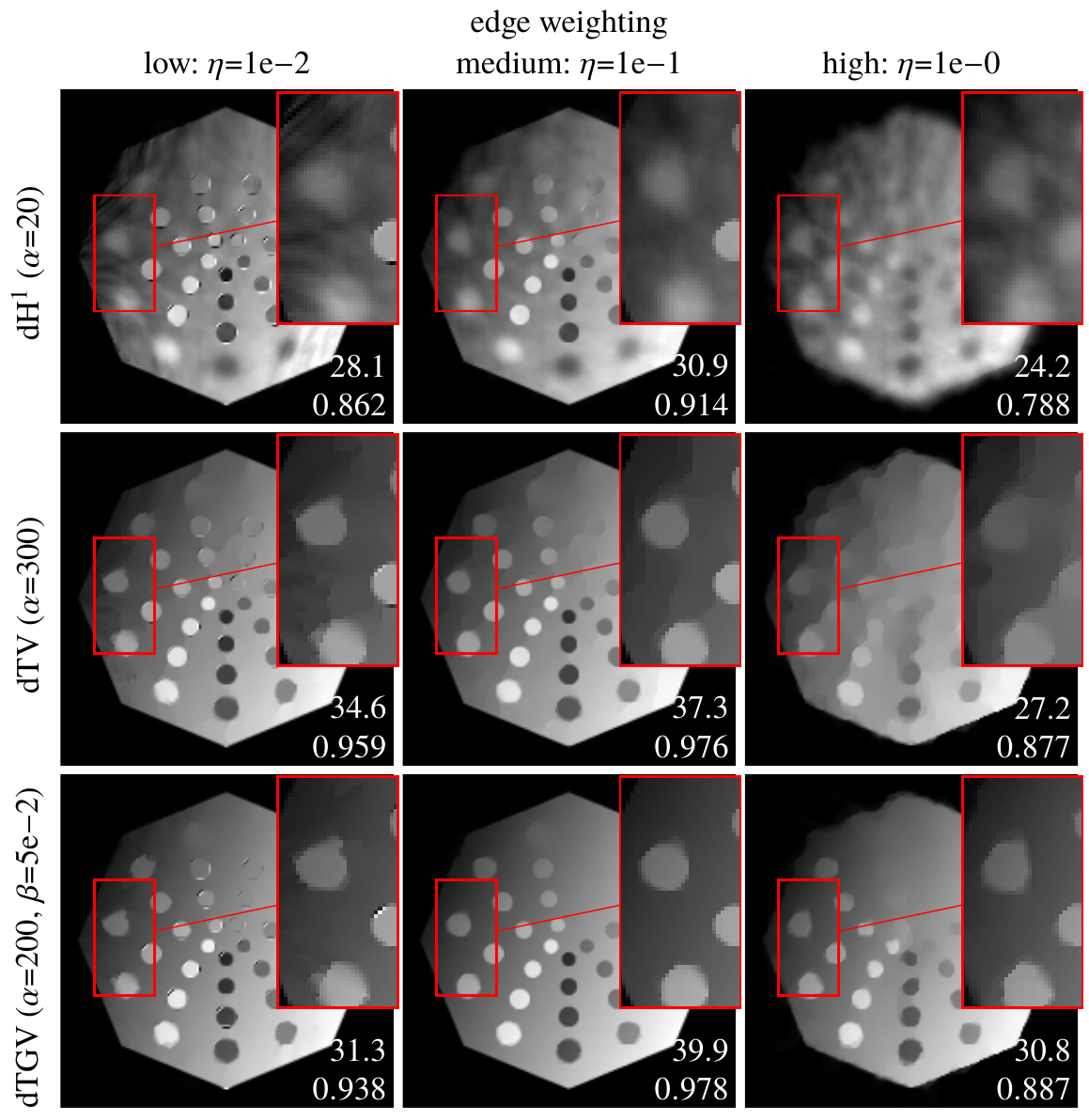}
\caption{Effect of edge weighting on directional models for test case \dataxray: increasing edge parameter $\eta$ from left to right. All other parameters where tuned to maximize the PSNR and visual image quality ($\gamma = 1$).} \label{fig:xray:edge:d}
\vspace*{4mm}
\includegraphics{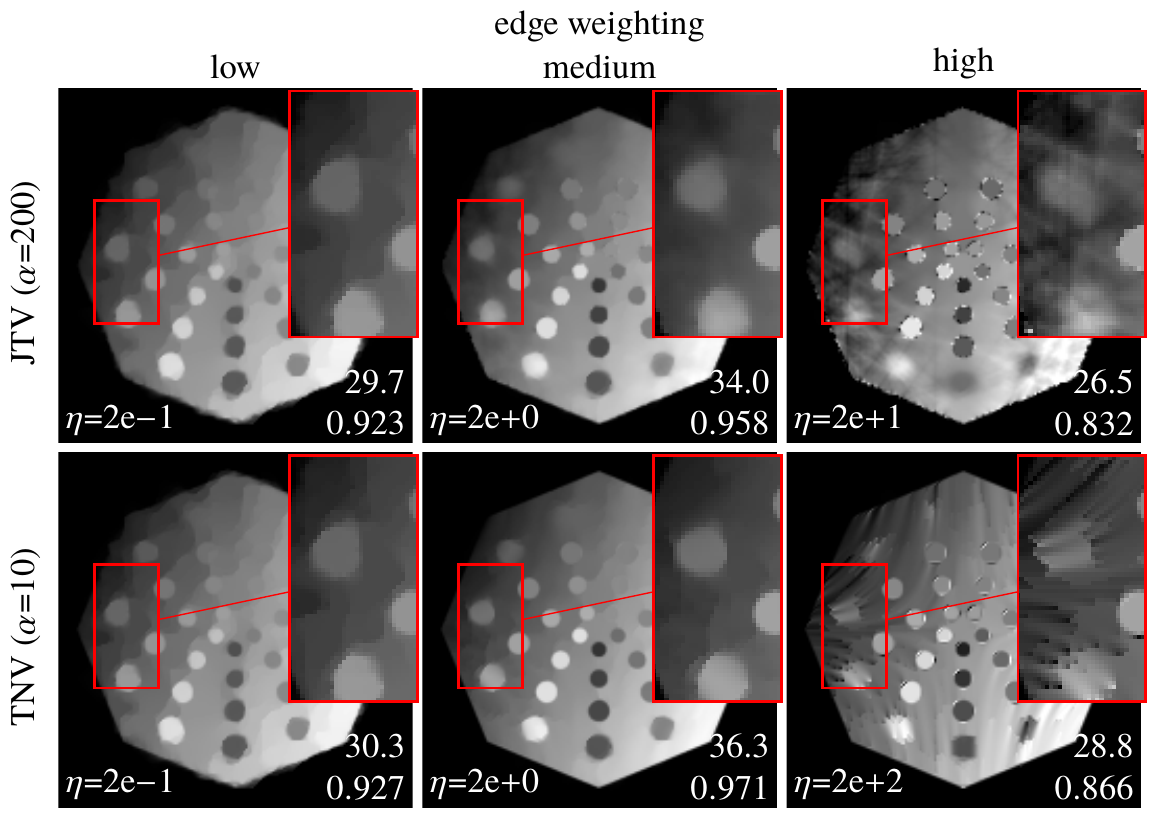}
\caption{Effect of edge weighting on joint total variation and total nuclear variation for test case \dataxray: increasing edge parameter $\eta$ from left to right. All other parameters where tuned to maximise the PSNR and visual image quality.} \label{fig:xray:edge:eta}
\end{figure}

All structure-promoting regularisers described in Section~\ref{sec:regularizers} have in common that they rely to some extend on the size of edges in the side information, i.e. $|\nabla v(x)|$. For $\JTV$ and $\TNV$ the actual values of $|\nabla v(x)|$ matter so that a parameter $\eta$ is needed to correct for this. For all other regularisers a parameter~$\eta$ is needed to decide which edges to trust and which not. The effect of this edge weighting parameter $\eta$ on all described regularisers is illustrated in Figures \ref{fig:xray:edge:w}, \ref{fig:xray:edge:d} and \ref{fig:xray:edge:eta}. The locally weighted regularisers (i.e. $\wHone$, $\wTV$ and $\wTGV$) and the directional regularisers (i.e. $\dHone$, $\dTV$ and $\dTGV$) have in common that if $\eta$ is too small, then small artefacts around the edges appear. This effect is more pronounced in locally weighted regularisers. If $\eta$ is too large, then the structure-promoting effect becomes too small. For joint total variation and total nuclear variation similar effects exist with reverse relationship to $\eta$.

\begin{figure}
\centering%
\includegraphics{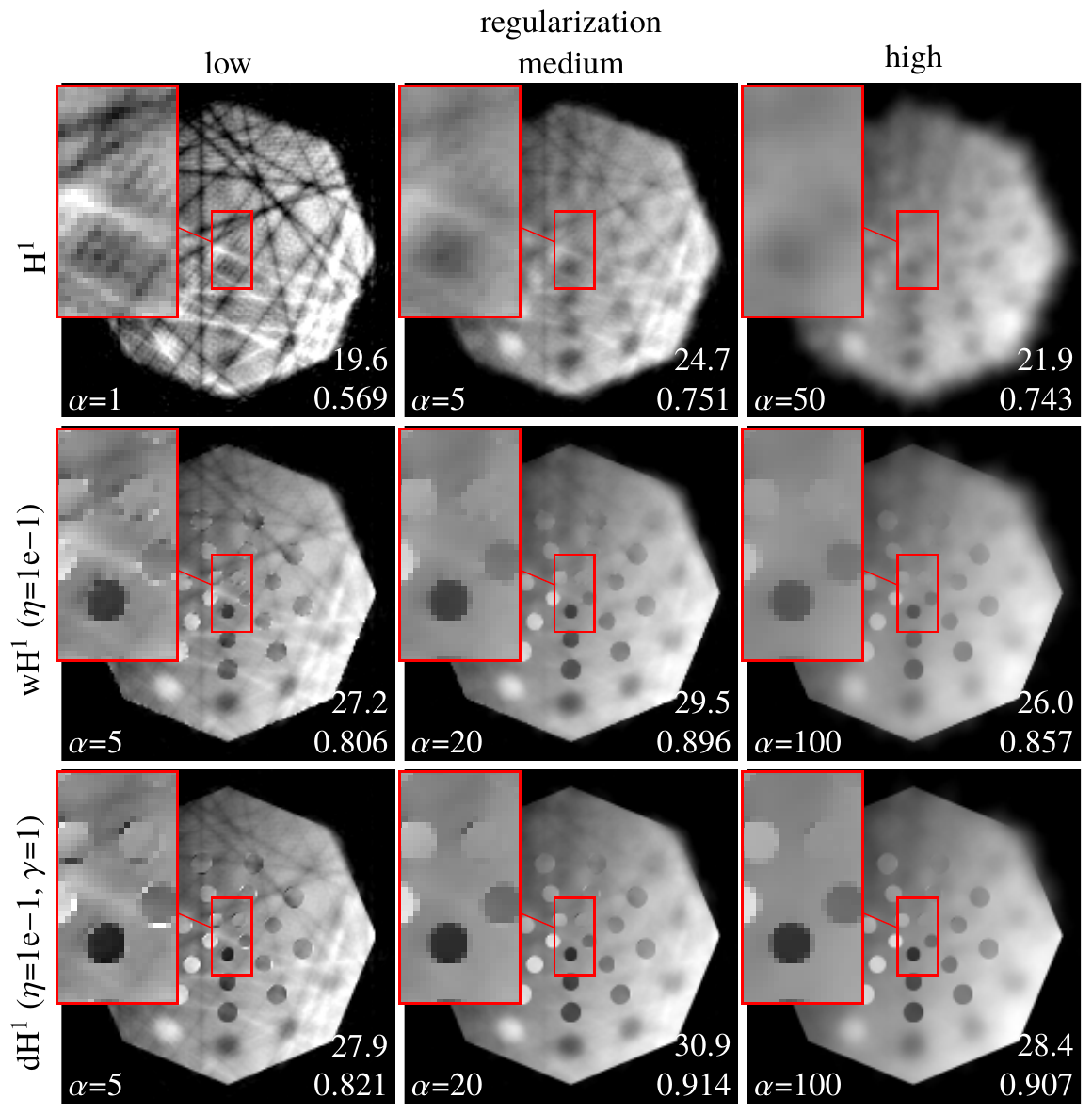}
\caption{\textbf{$H^1$-semi norm} based structure-promoting regularisers for test case \dataxray: increasing the regularisation parameter $\alpha$ from left to right. All other parameters where tuned to maximise the PSNR and visual image quality. All regularisers in this figure reduce to the $H^1$-semi norm in areas where the side information is flat.} \label{fig:xray:h1}
\end{figure}

\begin{figure}
\centering%
\includegraphics{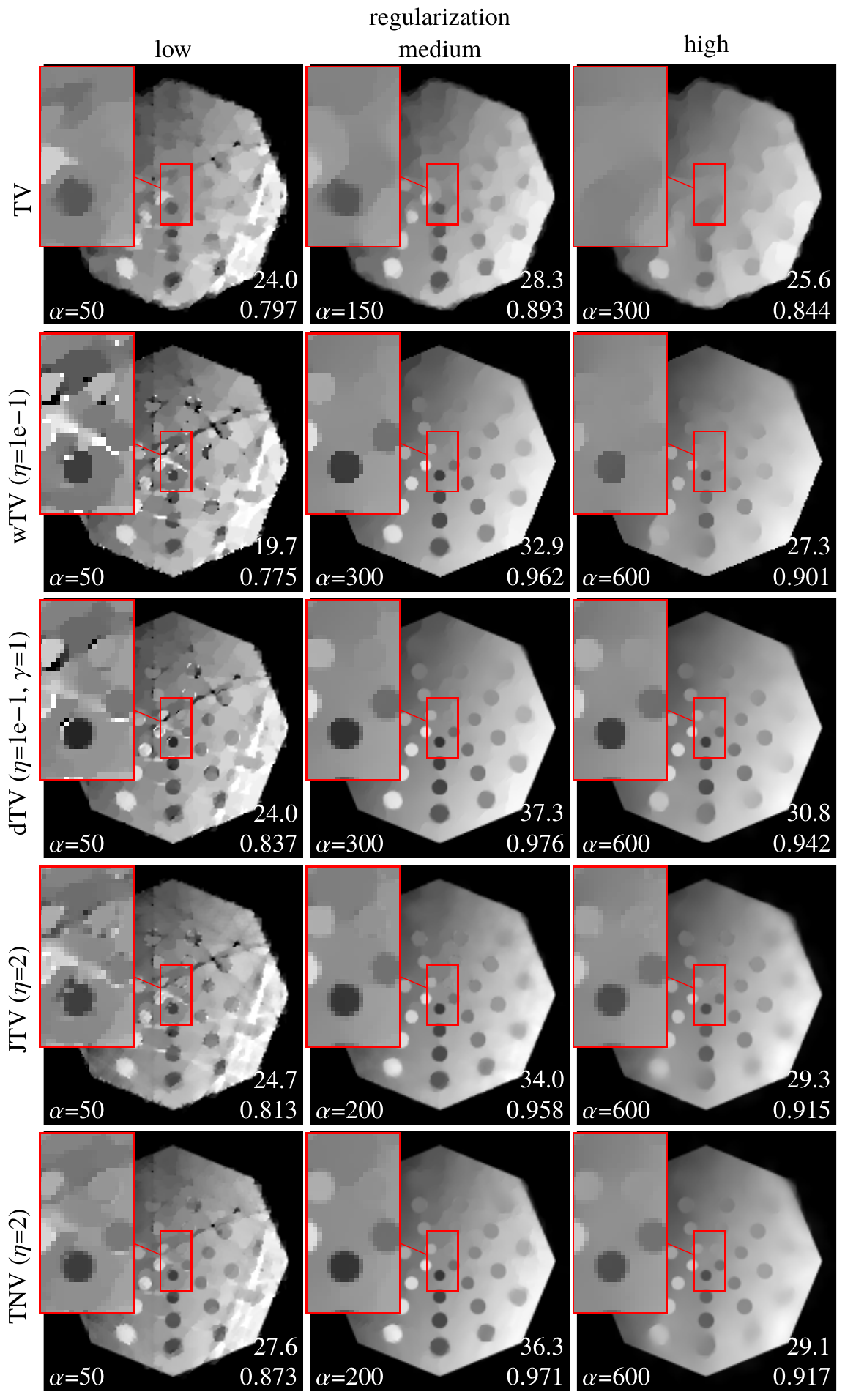}
\caption{\textbf{Total variation} based structure-promoting regularisers for test case \dataxray: increasing the regularisation parameter $\alpha$ from left to right. All other parameters where tuned to maximise the PSNR and visual image quality. All regularisers in this figure reduce to the total variation in areas where the side information is flat.} \label{fig:xray:tv}
\end{figure}

\begin{figure}
\centering%
\includegraphics{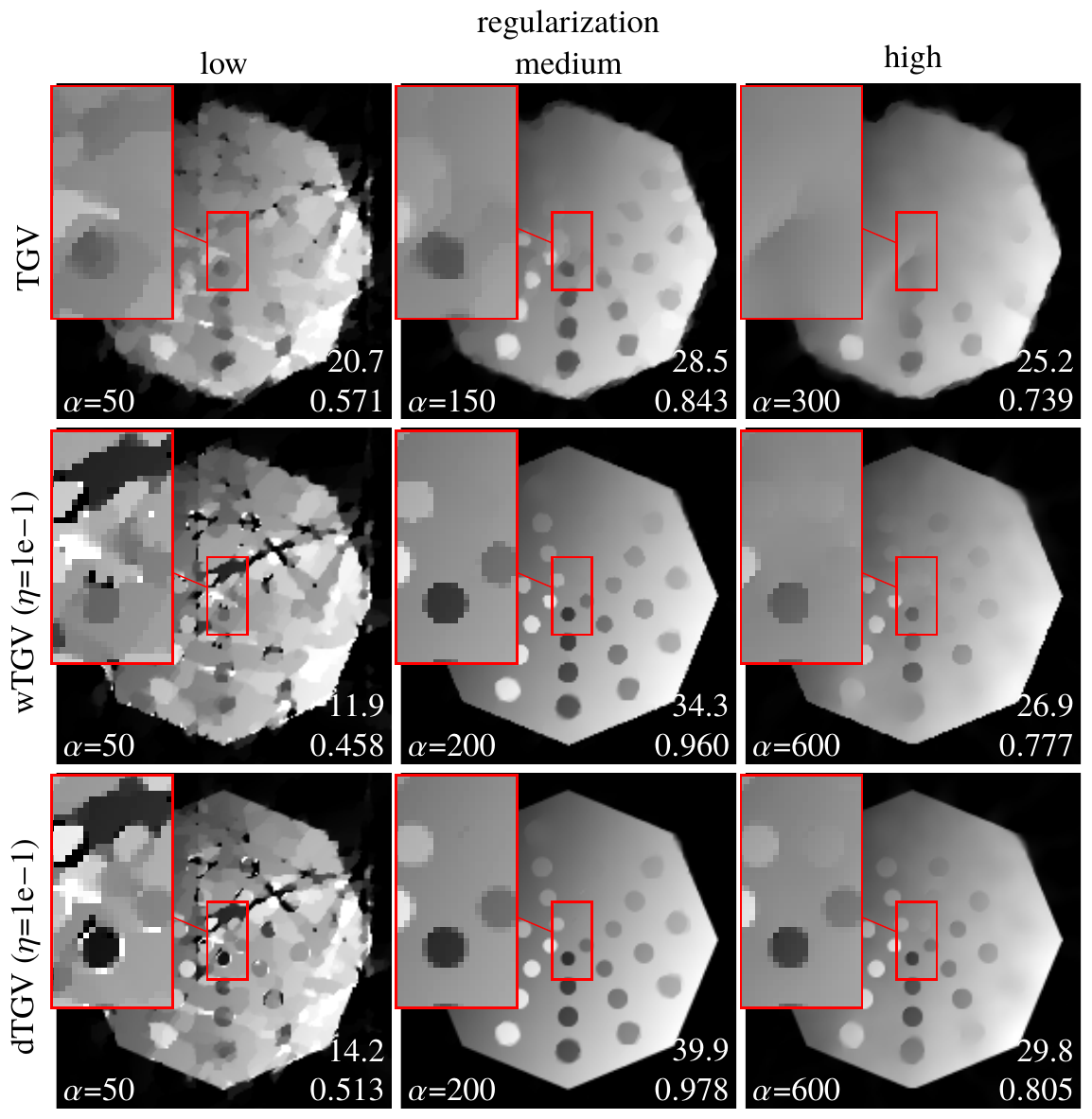}
\caption{\textbf{Total generalised variation} based structure-promoting regularisers for test case \dataxray: increasing the regularisation parameter $\alpha$ from left to right. All other parameters where tuned to maximise the PSNR and visual image quality ($\beta=\expnumber{5}{-2}$). All regularisers in this figure reduce to the total generalised variation in areas where the side information is flat.} \label{fig:xray:tgv}
\end{figure}

\runinhead{Effect of regularisation}
The effect of the regularisation parameter $\alpha$ on the solution is illustrated in Figures \ref{fig:xray:h1}, \ref{fig:xray:tv} and \ref{fig:xray:tgv}. All regularisers show the same behaviour if $\alpha$ is too small or too large. If the regularisation parameter is chosen too small then artefacts from inverting an ill-posed operator are introduced and if it is chosen too large then all regularisers oversmooth the solution. Note that all structure-promoting regularisers have an increased robustness in areas of shared structures.

\begin{figure}
\centering%
\includegraphics{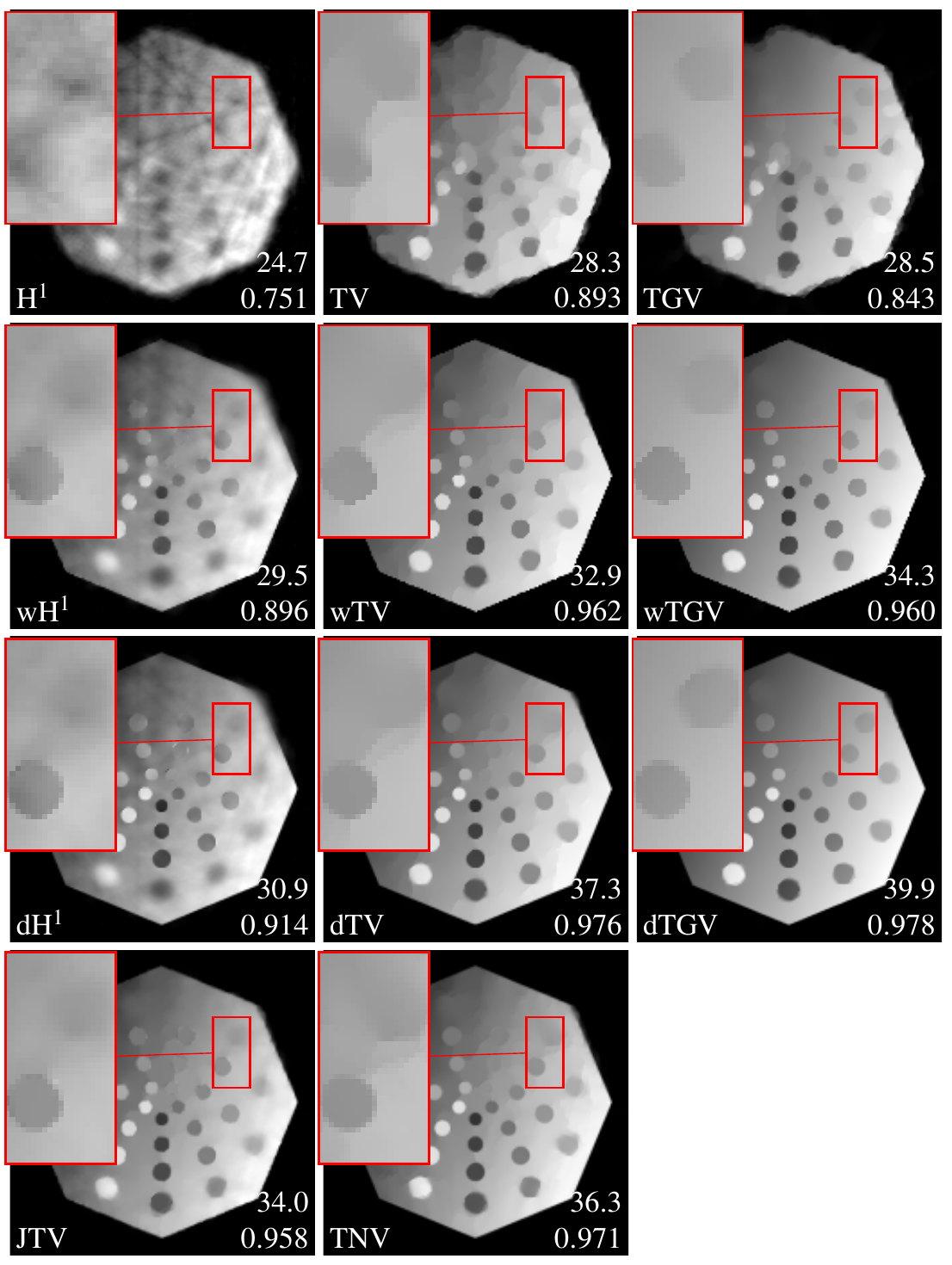}
\caption{Comparison of structure-promoting regularisers for test case \dataxray. All parameters where tuned to maximise the PSNR and visual image quality.} \label{fig:xray:comparison}
\end{figure}

\runinhead{Comparison of regularisers}
All eleven regularisers are compared in Figure \ref{fig:xray:comparison}. It can be seen that the structure-promoting regularisers perform much better in terms of PSNR and SSIM as their non-structure-promoting counterparts. Moreover, one can observe an interesting effect that the structure-promoting regularisers also perform visually better in regions where the structure is not shared, e.g. the outer ring of circles. This effect is most dominant for $\dTGV$ where the circle at the top left is clearly visible, while it is difficult to spot for many of the other regularisers.

\subsubsection{Test case \datasuper}

\begin{figure}
\centering%
\includegraphics{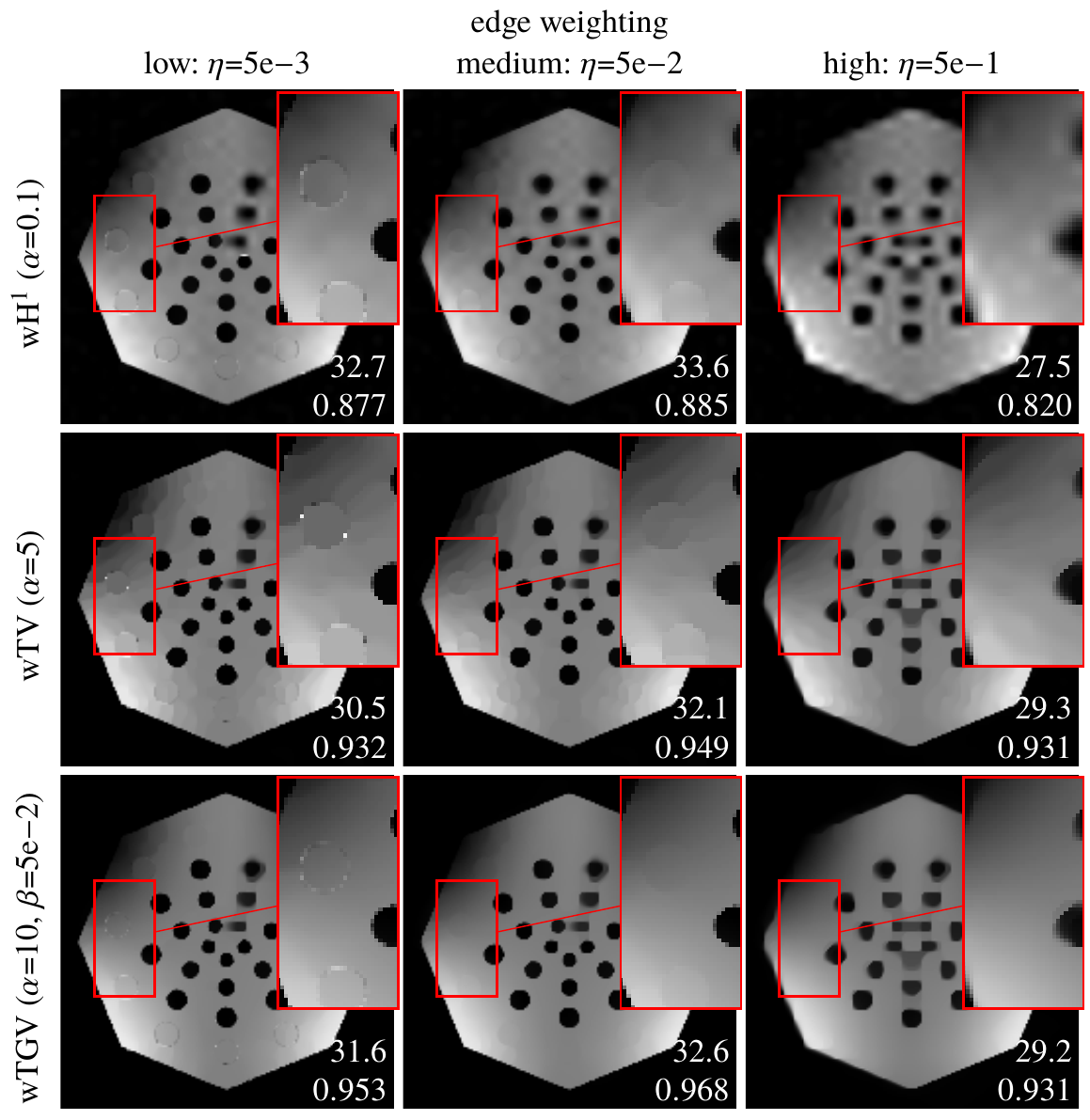}
\caption{Effect of edge weighting on locally weighted models for test case \datasuper: increasing edge parameter $\eta$ from left to right. All other parameters where tuned to maximise the PSNR and visual image quality.} \label{fig:super:edge:w}
\end{figure}

\begin{figure}
\centering%
\includegraphics{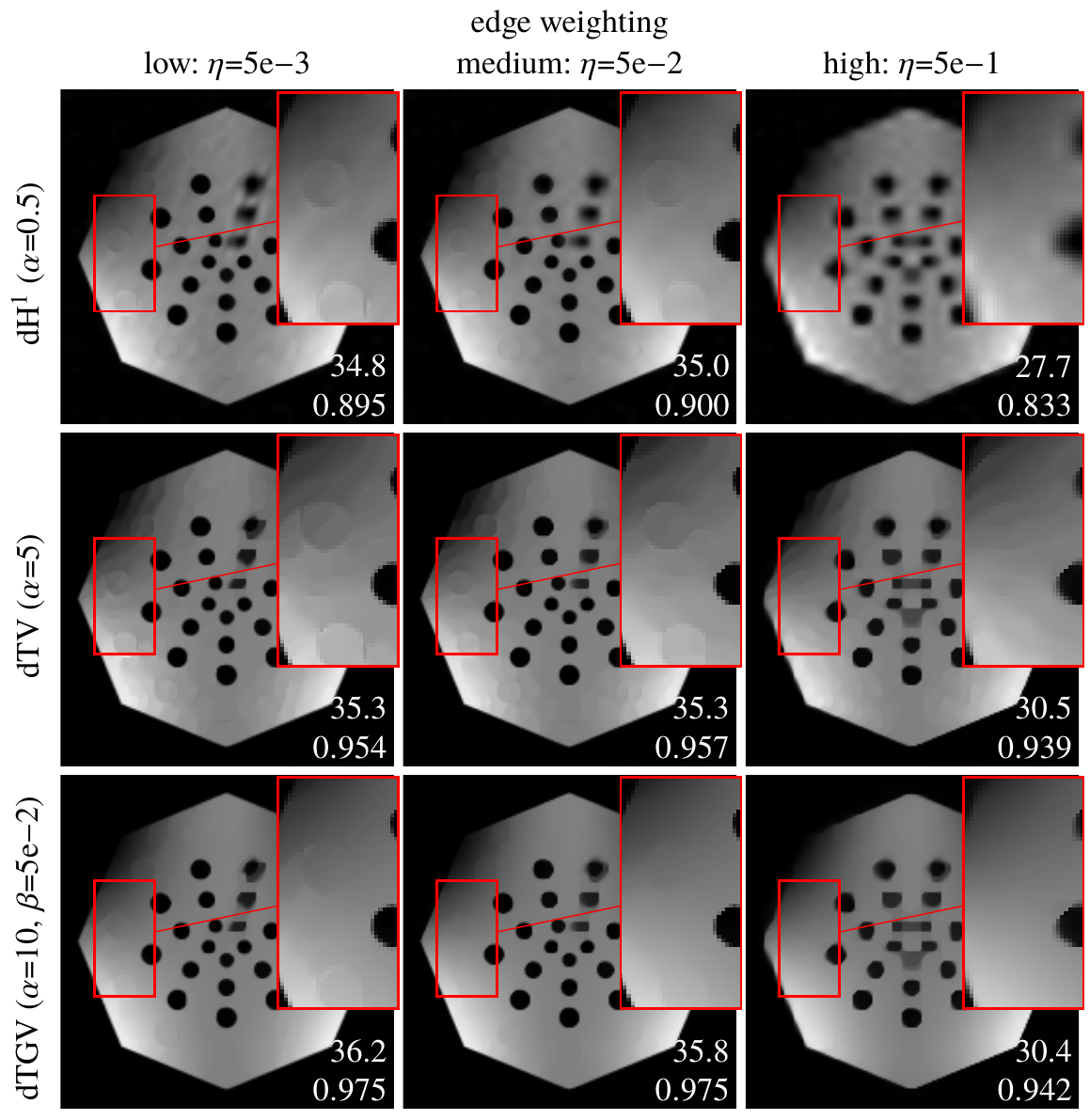}
\caption{Effect of edge weighting on directional models for test case \datasuper: increasing edge parameter $\eta$ from left to right. All other parameters where tuned to maximise the PSNR and visual image quality ($\gamma = 0.9$).} \label{fig:super:edge:d}
\vspace{4mm}
\includegraphics{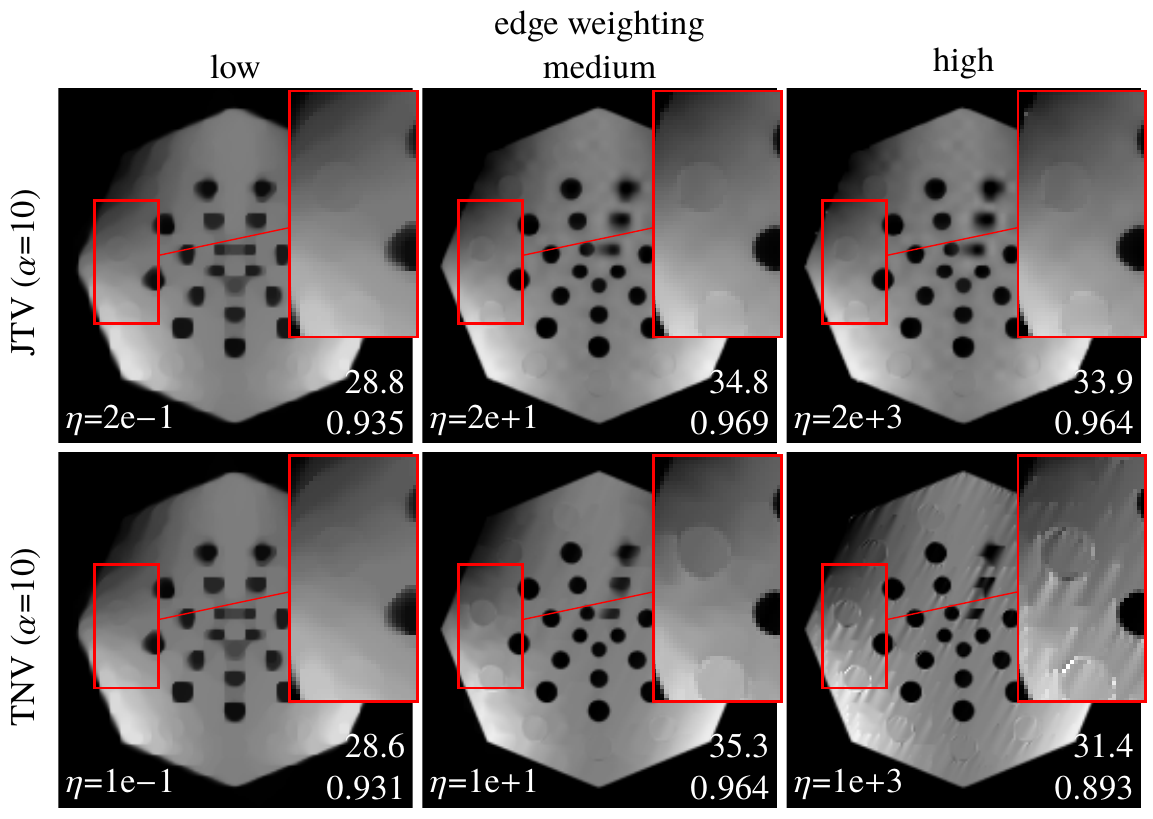}
\caption{Effect of edge weighting on joint total variation and total nuclear variation for test case \datasuper: increasing edge parameter $\eta$ from left to right. All other parameters where tuned to maximise the PSNR and visual image quality.} \label{fig:super:edge:eta}
\end{figure}

\runinhead{Effect of edge weighting}
Figures \ref{fig:super:edge:w}, \ref{fig:super:edge:d} and \ref{fig:super:edge:eta} show the effect of the edge weighting parameter $\eta$. One can make similar observations as in Figures \ref{fig:xray:edge:w}, \ref{fig:xray:edge:d} and \ref{fig:xray:edge:eta} for the test case \dataxray. In addition, one can observe from the close-ups that if $\eta$ is too small (or too large for $\JTV$ and $\TNV$), then ghosting artefacts may appear. Note that these are present for $\TNV$ even for a moderate choice of $\eta$.

\runinhead{Comparison of regularisers}

\begin{figure}
\centering%
\includegraphics{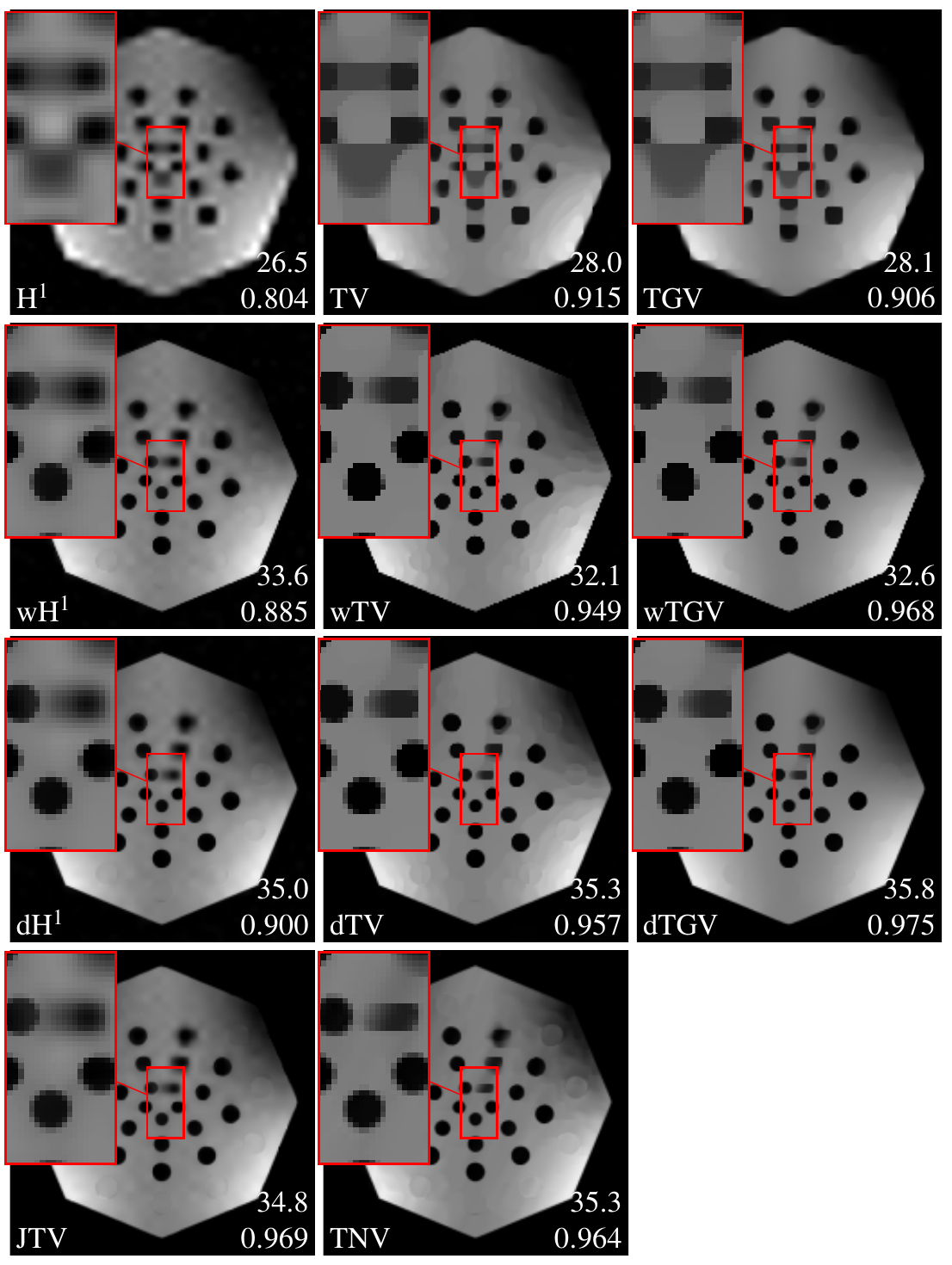}
\caption{Comparison of structure-promoting regularisers for test case \datasuper. All parameters where tuned to maximise the PSNR and visual image quality.} \label{fig:super:comparison}
\end{figure}

All regularisers are compared in Figure \ref{fig:super:comparison} for the test case \datasuper. It can be noted from all images that introducing structural information allows to resolve some of the inner circles which have been merged for regularisers which are not structure-promoting. Moreover, all total generalised variation based regularisers do not perform much better than the total variation based regularisers. The directional regularisers as well as $\JTV$ and $\TNV$ perform best in terms of PSNR for this example.

\subsection{Discussion on computational cost}
The median computing times for the numerical experiments are reported in Table \ref{tab:times}. The computing time of PDHG is mainly influenced by the dimensions of the models, the proximal operator and the forward model. As can be seen from the table, $\Hone$ and $\TV$ are roughly the same fast. $\TGV$ which uses a second primal variable in the space of the image gradient is significantly slower with about twice the computational cost. In all three cases, introducing isotropic weights (i.e. $\wHone$, $\wTV$ and $\wTGV$) increases the cost by about 6 seconds, and anisotropic weights (i.e. $\dHone$, $\dTV$ and $\dTGV$) by about 12 seconds. $\JTV$ is more costly than $\dTV$ but not as costly as $\TGV$. $\TNV$ is by far the most costly of all algorithms due to the need to compute singular value decompositions of $2 \times 2$-matrices for every pixel.

Since we run PDHG always for 3,000 iterations, we do not report computational time "till convergence" but computational cost for the full 3,000 iterations. It was observed at several occasions, see e.g. \cite{Ehrhardt2019pmb}, that including side information into the regulariser not only improves the reconstruction but also speeds up the algorithmic convergence. Intuitively this can be understood as more information is included into the optimisation problem.

Comparing the regularisers regarding their computational time versus image quality trade-off, it can be noted that $\TNV$ should not be chosen since it is not better than $\dTV$ at 7-10x the computational cost. Whether $\Hone$, $\TV$ or $\TGV$ based regulariser is desirable depends on each individual application. For each of them, there is a clear trend that one achieves better image quality by introducing more information, i.e. first isotropic information and then anisotropic information, each of which increases their computational cost. However, the increase in computational cost is so small that for most applications the directional variant is likely to be favoured.

\begin{table}
\caption{Computing times and PSNR for all tested regularisers.} \label{tab:times}
\begin{tabular}{p{1.2cm}P{2.5cm}P{2.5cm}P{2.5cm}P{2.5cm}}
\hline\noalign{\smallskip}
 & \multicolumn{2}{c}{computing time}  & \multicolumn{2}{c}{PSNR}  \\
regulariser & \hspace*{8mm}\dataxray\hspace*{8mm} & \datasuper & \hspace*{8mm}\dataxray\hspace*{8mm} & \datasuper \\
\noalign{\smallskip}\svhline\noalign{\smallskip}
$\Hone$ & \MySpace30.43 s & \MySpace18.34 s & 24.7 & 26.5 \\
$\wHone$ & \MySpace34.95 s & \MySpace22.31 s & 29.5 & 33.6 \\
$\dHone$ & \MySpace40.87 s & \MySpace27.80 s & 30.9 & 35.0 \\ \noalign{\smallskip}\hline\noalign{\smallskip} 
$\TV$ & \MySpace32.72 s & \MySpace18.63 s & 28.3 & 28.0 \\
$\wTV$ & \MySpace38.17 s & \MySpace22.05 s & 32.9 & 32.1 \\
$\dTV$ & \MySpace44.91 s & \MySpace29.48 s & 37.3 & 35.3 \\ \noalign{\smallskip}\hline\noalign{\smallskip} 
$\TGV$  & \MySpace71.33 s & \MySpace52.70 s & 28.5 & 28.1\\
$\wTGV$  & \MySpace77.67 s & \MySpace58.44 s & 34.3 & 32.6 \\
$\dTGV$ & \MySpace83.34 s & \MySpace61.65 s & 39.9 & 35.8 \\ \noalign{\smallskip}\hline\noalign{\smallskip}
$\JTV$ & \MySpace53.04 s & \MySpace39.05 s & 34.0 & 34.8 \\
$\TNV$ & 318.45 s & 290.42 s & 36.3 & 35.3 \\
\noalign{\smallskip}\hline\noalign{\smallskip}
\end{tabular}
\vspace*{-12pt}
\end{table}

\section{Conclusions and open problems}

This chapter introduced fundamental mathematical concepts on the structure of images and how structural similarity between images can be measured. The fundamental building blocks are the similarity based on edge sets and parallel level sets. These notions lead to several classes of structure-promoting regularisers all of which are convex and thereby lead to tractable optimisation problems when used in variational regularisation for linear inverse problems with convex data fits. While some of the regularisers are smooth and others are non-smooth, the resulting optimisation problem for all of them can be efficiently computed by PDHG. The effectiveness of these regularisers for the promotion of structure has been observed in many applications and was also illustrated in this chapter on two simulation studies.

\subsection{Open problems}
The mathematical framework for structure-promoting regularisers is by now well established and fairly mature. Open problems reside in practical problems in the translation of these techniques to applications which will also motivate further mathematical research.

\runinhead{Misregistration}
The biggest open problem is misregistration. All of the described regularisers assume that both images are perfectly aligned. Even in scanners which have two imaging modalities in the same system such as PET-MR, this assumptions is never perfectly fulfilled. This issue has not been addressed much in the literature. In~\cite{Tsai2018mic}, the authors proposed an alternating approach between image reconstruction and image registration with some success. In \cite{Bungert2018remotesensing}, the problem was formulated as a blind deconvolution problem so that translations can be compensated with a shifted kernel. The resulting optimisation algorithm is related to alternating gradient descent, thus alternates between incremental updates of each variable. This approach seems fairly efficient and robust but is limited to compensating translations and fails to find large deformations. The latter was approached heuristically in \cite{Bungert2018robust} but a generally applicable strategy is still to be found.

\runinhead{Extensions beyond two modalities}
It is natural to consider the case that more than one image is available as side information. For instance, in some remote sensing applications a colour photograph with high spatial resolution is available. Similarly in PET-MR, images of more than one MR sequence might be available. This setting has also been considered in \cite{Mehranian2017petplusmri} for a purely discrete model. Some of the regularisers to promote structural similarity in this chapter naturally extend to multiple images as side information, but this has not yet been properly investigated.

\runinhead{Applications}
As we illustrated in Section \ref{sec:intro}, there are many applications where structure-promoting regularisers were already used or are on the horizon. The list of potential target applications grows steadily with more and more multi-modality scanners being introduced. Next to the misregistration mentioned before, the biggest hurdle in real applications is the interpretation of images that were created by fusing information from several modalities. A common question is: "Which edges can I trust?" since often the reconstruction from multi-modality data would be performed on a finer resolution than for the single-modality case. For example, for PET-MR the reconstruction of PET data with an already reconstructed MR image as side information can be performed on the native MRI resolution. The answer might be that such an image should not be interpreted as a PET image, but in fact as a synergistic PET-MR image.

\runinhead{Joint reconstruction}
Throughout this chapter the focus was on improving the reconstruction of one image with the aid of another modality used as side information. Since the other image is rarely acquired directly, it is natural to aim to reconstruct both images simultaneously rather than sequentially. While conceptually appealing this strategy leads to many more complications than the approach discussed in this chapter which is sometimes referred to as one-sided reconstruction. While the mathematical framework for one-sided reconstruction is quite mature, the framework for joint reconstruction is despite a lot of research effort in the last 10 years still in its infancy. Fundamental problems like computationally tractable and efficient coupling of modalities is still unsolved. The appealing strategy of making use of the solid mathematical foundations of one-sided reconstruction for joint reconstruction in a mathematical sound and computationally tractable way is still not possible to date.

\section*{Acknowledgements}
The author acknowledges support from the EPSRC grant EP/S026045/1 and the Faraday Institution EP/T007745/1. Moreover, the author is grateful to all his collaborators which indirectly contributed to this chapter over the last couple of years.

\end{document}